\documentclass{ar-1col}
\usepackage[dvipsnames]{xcolor}
\usepackage{url}
\usepackage{gensymb,amssymb}
\usepackage{color,graphicx}
\usepackage{geometry}
\usepackage{enumitem}
\usepackage{comment}
\usepackage{natbib,xcolor,color}
\usepackage{colortbl}

\newcommand\simlt{\lower.5ex\hbox{$\; \buildrel < \over \sim \;$}}
\newcommand\simgt{\lower.5ex\hbox{$\; \buildrel > \over \sim \;$}}

\definecolor{orange}{RGB}{170, 170, 170}
\definecolor{red}{RGB}{250, 150, 170}
\definecolor{green}{RGB}{170, 200, 250}

\jname{Xxxx. Xxx. Xxx. Xxx.}
\jvol{AA}
\jyear{YYYY}
\doi{10.1146/((please add article doi))}

\begin{document}

\markboth{No\'emie Globus and Roger Blandford}{Ultra High Energy Cosmic Rays}

\title{Ultra High Energy Cosmic Rays}
\author{No\'emie Globus$^{1,2,3}$ and Roger Blandford$^{2}$
\affil{$^1$Instituto de Astronomía, Universidad Nacional Autónoma de México Campus Ensenada, A.P. 106, Ensenada, BC 22860, México}
\affil{$^2$Kavli Institute for Particle Astrophysics and Cosmology, Stanford University, Stanford, CA 94305, USA}
\affil{$^3$Astrophysical Big Bang Laboratory, RIKEN, Wako, Saitama, Japan}
}

\begin{abstract}
Ultra High Energy Cosmic Rays, UHECR, are  charged particles with energies between $\sim10^{18}\,{\rm eV}$ and $\sim3\times10^{20}\,{\rm eV}\sim50\,{\rm J}$.  They exhibit fundamental physics at energies inaccessible to terrestrial accelerators, challenge experimental physics and connect strongly to astronomical observations through electromagnetic, neutrino and even gravitational wave channels. There has been much theoretical and observational progress in the sixty years that have elapsed since the discovery of UHECR, to divine their nature and identify their sources.
\begin{itemize}
\item{The highest energy UHECR appear to be heavy nuclei\\ with rigidity extending up to $\sim10\,{\rm EV}$;}
\item{A significant ($6.9\sigma$) dipole anisotropy has been measured but our\\ poor understanding  of the Galactic magnetic fields makes this hard\\ to interpret;}
\item{The UHECR luminosity density is $\sim 10^{44}$ erg Mpc$^{-3}$ yr$^{-1}$\\  
which constrains explanations of their origin;}
\item{The most promising acceleration mechanisms involve diffusive\\ shock acceleration and unipolar induction;}
\item{The most promising sources include intergalactic accretion shocks,\\ and relativistic jets from stellar-mass or supermassive black holes.}
\end{itemize}
We explore the prospects for using the highest energy events, combined with multimessenger astronomy, to help us solve the riddle of UHECR.
\end{abstract}

\maketitle
\tableofcontents

\section{INTRODUCTION}
\begin{extract}
``Contemporary astrophysics is faced by a number of acute problems. One of them concerns dark matter, which one might (perhaps mischievously) qualify as the study of particles which should exist... but until farther notice, don’t. Ultra high energy cosmic rays constitute the inverse problem: particles which do exist... but perhaps shouldn’t.'' -- Ludwik M. Celnikier 
\end{extract}
Ultra High Energy Cosmic Rays, UHECR - conventionally defined to be cosmic rays with energies above 1 EeV\footnote{UHECR bring together communities that use different units. We will adopt the practices of those most directly involved with the measurements and provide occasional translations where this may be helpful.} - 
\begin{marginnote}[-2pt]\entry{Cosmic Ray Energy}{This is what is  directly measured and denoted as $E$. It is usually expressed in eV with MeV, GeV, TeV, PeV, EeV, ZeV corresponding to $10^{6,9,12,15,18,21}\,{\rm eV}$. UHECR have  $E>1\,{\rm EeV}\equiv0.16\,{\rm J}$.}\end{marginnote} 
are the most energetic subatomic particles observed in nature. At 10 EeV, UHECR are only detected at a rate of about 1 per km$^2$ per year.  
When an UHECR interacts with the atoms of our earth's atmosphere, it triggers the formation of a very large number of secondary particles \citep{Engel2011}\footnote{In preparing this review, we have been confronted with an order of magnitude more papers than we can cite. We have tried to give preference to those that convey the most basic understanding and provide a good path to the current literature. Where we have failed, we apologize.}, with correlated acoustic \citep{1985ICRC....8..333G}, radio \citep{2017PrPNP..93....1S}, ultraviolet Cherenkov \citep{2011NuPhS.212...13W} and ultraviolet fluorescence \citep{1967PhDT........28B} emissions.\begin{marginnote}[-2pt]\entry{Extensive Air Shower, EAS}{The production  of billions of secondary, lower energy particles by a primary UHECR when it hits the top of the atmosphere.}\end{marginnote}The atmosphere, with a column density of $\sim1030\,{\rm g\,cm}^{-2}$,  acts as a calorimeter, absorbing most of the energy of the primary particle and depositing mostly secondary muons \begin{marginnote}[-2pt]\entry{Muons}{Discovered in 1936 by Anderson and Neddermeyer in cosmic ray showers, muons are elementary particles with a lifetime of 2.2~$\mu$s and $\sim$207 times the mass of an electron. }\end{marginnote} on the ground. The first observation of secondary cosmic-ray muons dates back to the 1940s, when ionization chambers were used as detectors. Over time, the detection method migrated to  Geiger-Müller counters, and then, to plastic scintillators and  Cherenkov water tanks that are used  today \citep{2019EPJWC.21000001W}.  Ground particle detectors associated with fluorescence telescopes  catch the  secondary charged particles and the associated radiation. The data is then compared to shower models to   deduce the energy, mass and direction of the primary particle.


 The year 2023 marked the 60$^{\rm th}$ year anniversary of the very first paper on  UHECR, reporting the discovery of a primary cosmic ray with energy $\sim100$~EeV \citep{PhysRevLett.10.146}. Cosmic rays with energies beyond 100 EeV are commonly referred to as Extreme Energy Cosmic Rays, EECRs, and are very rare. Only an handful of  events with similar energies  have been detected since then. A particle in this energy range, on average, hits a square kilometre of Earth less than once per century and require  experiments with a large geometrical exposure and decades of observational effort to accumulate sufficient statistics.  It took close to 50 years to find the end of the remarkable, power-law energy spectrum  which was confirmed independently by  UHECR experiments in the northern and southern hemisphere  \citep{2008PhRvL.100j1101A, 2008ICRC....4..335Y}. The highest energy validated is $\sim300$~EeV (near 50 J per nucleus, corresponding to a 250 g baseball traveling at 72 km h$^{-1}$)\footnote{but having the momentum of a small snail!}. 

The arrival direction of UHECR on Earth do not point to their sources, because UHECR  are charged particles which are deflected by cosmic magnetic fields. The relevant parameter for the motion of charged particles in magnetic fields is their Larmor radius (or gyroradius), $r_L\propto R/B$, where {\it R} is the particle rigidity.\begin{marginnote}[-2pt]\entry{Rigidity}{This is momentum per unit charge and connects most directly to the physics of the particle acceleration and propagation. This is denoted as $R=E/Ze$ for ultra-relativistic particles and is measured in volt.}\end{marginnote}Identifying the nature of the primary cosmic rays is therefore necessary to model their trajectories in Galactic and extragalactic magnetic fields and reconstruct the direction of their sources. Unfortunately,  the composition of  UHECR is not well known, because the modeling of the EAS at these high energies requires an extrapolation of the physics of particle interactions obtained at the Large Hadron Collider, LHC.  
Therefore, the origin of UHECR has been an open question for over half a century. 

The goal of this review is to describe the current observational status and source candidates of UHECRs,
in the light and shadow of our present understanding, as well as their nature, acceleration, and transport in
magnetic fields. To introduce the subject further, let's unfold a triptych presenting UHECR from three different perpectives: the particle physicist's view, the cosmic-ray experimentalist's view,  and the astrophysicist view, so as to address three  questions at the heart of UHECR physics: What are they? How do we detect them? What is their origin?.

\begin{textbox}[ht]
\section{CENTER OF MASS ENERGY}
The center-of-mass energy $\sqrt{s}$ is  the largest total mass that can be produced in a collision process. For two particles $A$ and $B$ with rest-mass $m_A$ and $m_B$, $s =  m_A^2 c^4 + 2E_{A,{\rm lab}}m_B c^2 + m_B^2 c^4$ where $E_{A,{\rm lab}}$ is the energy of $A$ in the laboratory frame and $B$ is at rest.
\end{textbox}

\subsection{Through the particle physicist's eyes: ``What are they?''}\label{partphys}
To the particle physicist, UHECR exhibit collisions with center of mass energies ({\bf Sidebar Center of Mass Energy}) $\sqrt{s}$, a hundred times those attainable at the LHC\footnote{\citet{hut83} presented an interesting argument that collisions between pairs of energetic cosmic rays  on our past light cone, whose consequences could have been experienced, had enormous center of mass energies and yet had not ``collapsed the vacuum''. Consequently, we had nothing to fear from the LHC!} and the ever opportunity to uncover new physics.  Upon impact with earth atmosphere, cosmic rays produce showers of secondary particles called Extensive Air Showers, EAS. Every EAS constitute a unique multimessenger signature of hadronic interactions ({\bf Sidebar Hadronic interaction}), which depends on the energy and nature of the primary particle. Each of these interactions give rise to new particles and are the foundation of EAS modeling. A correct understanding of the physics of the EAS is essential to determine the nature  of the primary UHECR. However, the interactions of UHECR in the atmosphere are not yet fully understood. We review below the current knowledge of the theory and models of EAS.

\begin{figure*}
\includegraphics[width=5.55in]{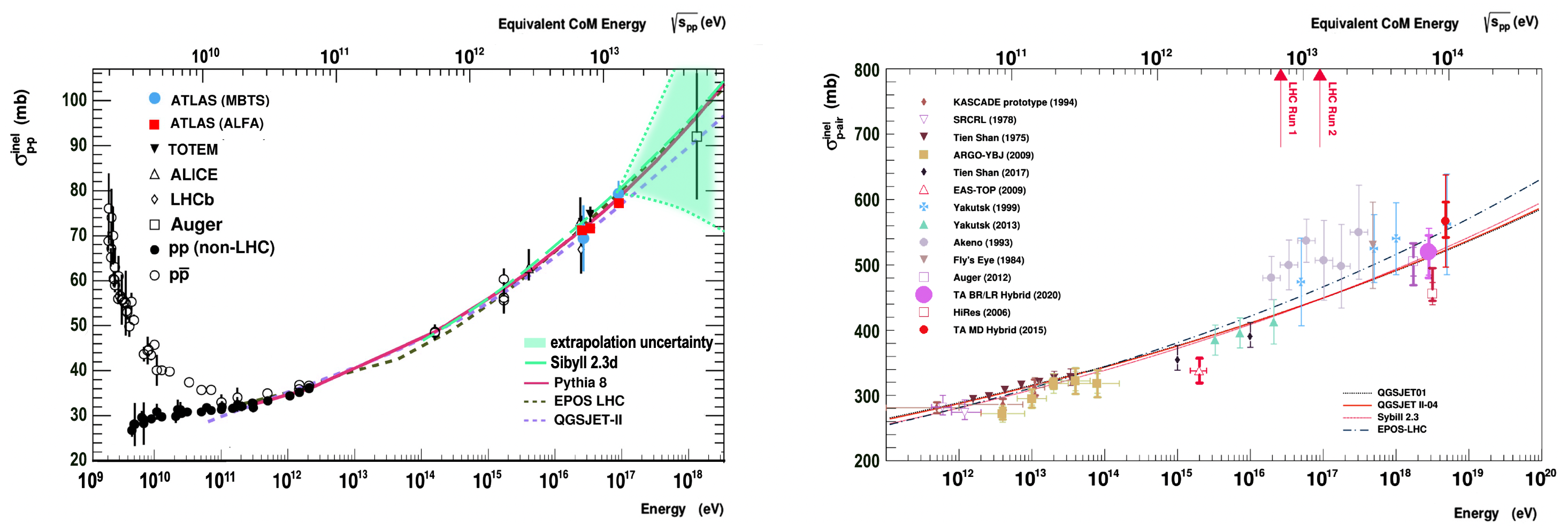}
\caption{Left:  Inelastic cross-section measurements and model predictions for proton-proton interactions as a function of the energy. Figure adapted from \citep{2023EPJC...83..441A}. Right: Same for proton-air interactions.  Adapted from \citep{2020PhRvD.102f2004A}}
\label{fig1}
\end{figure*}

The development of the EAS in the atmosphere can be understood through the modeling of the inelastic production of secondary particles: pions, kaons, muons, baryon-antibaryon pairs and other particles, but also leptons and photons  which are produced through bremsstrahlung, Cherenkov and pair production  mechanisms. The underlying theory that describes hadronic interactions is quantum chromodynamics \citep[e.g.][for a recent review]{Gross23} where the fundamental fields are quarks and gluons. In the limit of large momentum transfer between these fundamental fields, one can use perturbation theory to compute interaction cross sections. The perturbation theory fails in the elastic limit of these processes and, since quantum chromodynamics cannot yet be solved in the nonperturbative regime, experiments are necessary. The highest collision energy reached so far at the LHC is $\sqrt{s} \approx13$~TeV  while UHECR have  $0.2\lesssim\sqrt{s}\lesssim3$~PeV. Since the secondary nuclear production cross sections cannot  be measured nor calculated,  UHECR physicists rely on extrapolations of the existing hadronic interaction models,   QGSJet-II-04 \citep{2011PhRvD..83a4018O},   EPOS-LHC \citep{2015PhRvC..92c4906P},  SIBYLL-2.3 \citep{Riehn:2015oba} and DPMJET \citep{Fedynitch15} to predict the evolution of the cross sections with energy.   The energy evolution of the inelastic cross-sections for proton-proton ({\it p-p}) and proton-air ({\it p-air}) interactions derived from recent measurements is shown in Fig.~\ref{fig1}.

\begin{textbox}[ht]
\section{HADRONIC INTERACTION}
Hadronic interaction refers to the strong interaction of a hadron with a nucleus. The simplest example of a hadronic interaction is $p + p/N \rightarrow \pi + X$ production, where $p$ is a proton, $N$ a nucleus, and $X$ can be any kind of secondary particle produced during the interaction. The most frequently produced secondary hadrons are charged and neutral pions. Charged kaons with a slightly shorter lifetime require higher energies. The inelastic cross section for  proton-proton $\sigma_{pp}$ and proton-air $\sigma_{p-air}$  interactions is shown in Fig.~\ref{fig1}. Above $E_{\rm lab} = 3$~GeV, $\sigma_{p-p}$ is given approximately by the expression $\sigma_{p-p}(E_{\rm lab})=30.364-1.716\log(E_{\rm lab}) +0.981 \log(E_{\rm lab})^2\,\,{\rm mb}$ (EPOS-LHC model). The inelasticity --- the fractional energy change in a collision --- is $\kappa\approx0.5$  at $\sqrt{s}\approx3$~GeV. Its variation with energy depends on the hadronic interaction model employed \citep{Pierog:2017nes}. 
\end{textbox}

\vspace{2pt}
Many aspects of the relation between hadron production and shower observables can be qualitatively understood within the Heitler model \citep{Heitler:1936jqw} and its Heitler-Matthews extension \citep{Matthews:2005sd}. Every particle undergoes a splitting after it travels a fixed distance related to the interaction length.  
It is assumed that every hadronic interaction results in $\sim15$ pions, 2/3 of which are charged  and 1/3 are neutral.
The neutral pions  immediately decay into two photons ($\pi^0\rightarrow 2\gamma$)  which will interact with the atoms of the atmosphere to produce a pair of electron-positron 
in approximately one interaction length. These daughter electrons and positrons will also interact with air on a similar distance scale (also called radiation length) to create bremsstrahlung photons, again dividing the energy. The charged pions  interact again to create new pions, which decay into muons and neutrinos once their energy is below the threshold for producing new pions ($\pi^+\rightarrow \mu^++\nu_\mu$, $\pi^-\rightarrow \mu^-+\bar{\nu}_\mu$). 

The EAS are  usually  described in a simplified way as the composition of three components: a hadronic (nuclei fragments, protons, neutrons) cascade, an electromagnetic cascade (gamma-rays and electron-positron pairs), produced by the neutral pions, and a muonic cascade, produced by the charged pions \citep{Engel2011}, as illustrated in Fig.~\ref{fig2}. The hadronic cascade contains the long-lived secondary hadrons forming the  shower core.  The growth of the electromagnetic cascade is governed by bremsstrahlung and pair production until the energy of the photons fall below the threshold energy for producing new pairs. Energetic muons, of which 90\% are produced in the hadronic cascade due to the decay of pions and kaons, propagate through the atmosphere with small energy losses and reach the ground, with decay lengths $\sim6(E_\mu/1\,{\rm GeV})\,{\rm km}$, almost unattenuated. 
 
The LHC-tuned hadronic event generators do not correctly reproduce in detail the EAS observables  that can be probed by UHECR detectors \citep{2016PhRvL.117s2001A,2018PhRvD..98b2002A}. In particular,  EAS simulations  show a significant muon deficit with respect to the measurements for all hadronic interaction models. This is called the ``Muon Puzzle'' \citep{2022Ap&SS.367...27A}. An effort is made to upgrade existing experiments  and design the next-generation  to extend our understanding of hadronic interactions beyond $\sqrt{s}=100$~TeV \citep{2019EPJWC.21002004D, 2023NatSR..1316091K, Tiberio23, 2024EPJC...84..696K}. “As
of April 2025, we still do not have a clear answer to the question {\it "What are they?"} but we do know that UHECR  are neither  dominated by photons \citep{2009APh....31..399P} nor neutrinos \citep{2008ApJ...684..790A,2009PhRvD..79j2001A}, and that a pure proton composition is disfavored at the highest energies \citep{2024PhRvD.109j2001A}. 

\subsection{Through the cosmic-ray physicist's eyes: ``How do we detect them?''}\label{detection}
\begin{figure}[h]
\includegraphics[width=5.5in]{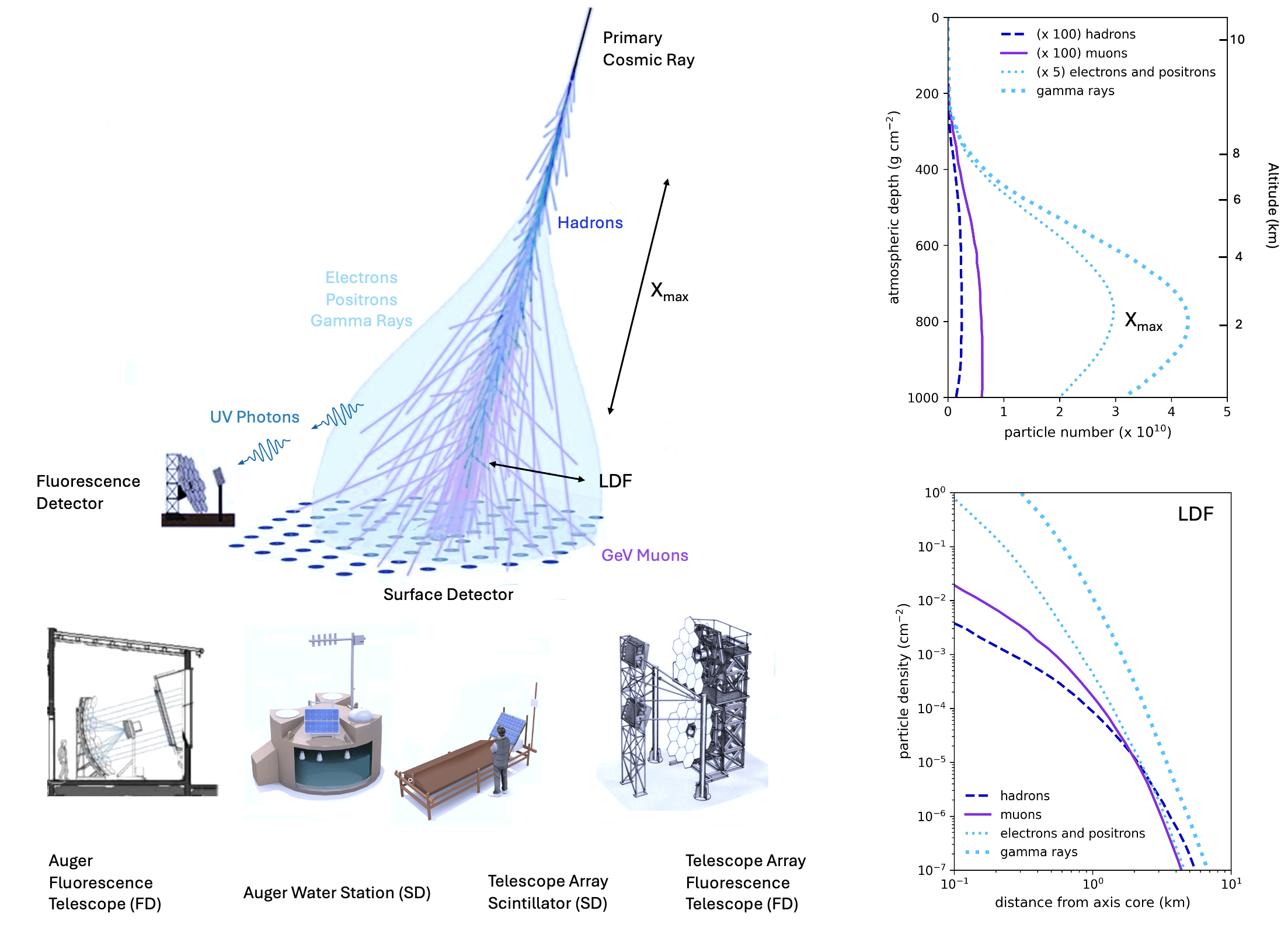}
\caption{The illustration \citep[adapted from ][]{Coleman2023} shows the development of an EAS in the atmosphere and the different surface detectors, SD, used by TA and Auger. The sky over the  SD is viewed by fluorescence detectors, FD, which are used for the calorimetric energy determination of an EAS from the energy deposited in the atmosphere during its development.  The EAS can be divided into three components: the hadronic; the muonic (muons and neutrinos), and the electromagnetic (electrons, positrons and gamma rays). The hadronic component comes from the interaction of UHECR or secondary particles with the molecules of the atmosphere and is mostly mesons like pions and kaons which will then decay into muons and neutrinos. The electromagntic component derives from subsequent cascades involving bremsstrahlung and pair creation until the energy falls below threshold for creating new pairs, which occurs at $X_{\rm max}$. A vertical shower of a proton with 10 EeV, at ground, has $\approx10^{11}$ secondary particles with energy above 90~keV and a shower depth of $\approx$10 km. The plots show the average longitudinal (upper panel) and lateral (lower panel) shower profiles of the hadronic, muonic and electromagnetic components generated with the CORSIKA code \citep[adapted from][]{2018pmabook} for a 10 EeV proton primary.  Detectors illustrations credits: Pierre Auger Observatory and Telescope Array.} 
\label{fig2}
\end{figure}

To the traditional cosmic ray physicist, UHECR represent the culmination of a century's development of experimental technique with the achievements of the two most advanced  cosmic-ray observatories, the Pierre Auger Observatory \citep[][]{PierreAuger:2004naf}, hereafter Auger, and the Telescope Array experiment \citep[][]{2008NuPhS.175..221K}, hereafter TA. Both experiments are based on what is called a ``hybrid'' observational approach,  combining the observation of secondary particles that reach the ground (mostly muons) detected by the surface detectors, SD, and the observation of the fluorescence emission - produced by the EAS when it excites the molecules in the atmosphere - detected by the fluorescence detector, FD, as illustrated in Fig.~\ref{fig2}. The goal is to use the FD and the SD together to measure the properties of the primary particle: energy, mass  and arrival direction. The spatial and temporal structure of the EAS depends in different ways on the mass and energy of the primary particle. Therefore, at least two  independent measurements of the  energy and the mass are necessary to determine the properties of the primary particle. The EAS observables are:

$\bullet$ {\it The  Longitudinal Development ($X$)}: 
The number of secondary particles versus the slant depth of atmosphere traversed (conventionally measured in grams per square centimeter). The point where the number of particles is greatest defines the  depth of the shower maximum, $X_{\rm max}$.

$\bullet$ {\it The  Lateral Distribution Function (LDF)}: the particle density at ground vs. distance to the impact point (shower core). The detected particles can be muons or electromagnetic component (electrons, positrons and gamma-rays) or a mixture of all particles, depending on the detectors.

$\bullet${\it The associated light emission}: the beamed Cherenkov emission, which can be detected by looking into the core of the shower, and the isotropic fluorescence  and radio emissions.
\smallskip

\begin{textbox}[ht]
\section{SURFACE DETECTORS} 
For TA, the SD array is made of plastic scintillators which provide a measure of the energy loss by a particle that goes through a medium (also called the {\it stopping power}). The working principle of the scintillator is to detect the  brief flash of light produced when an ionizing particle  goes through a  material. The light is measured by a photomultiplier tube that produces a voltage pulse when the light falls on it. The intensity of the light is a direct measure of the stopping power converted into visible light by the photomultiplier.

For Auger, the SD array is made of water tanks. When a particle moves through a medium at a velocity greater than that of the light in that medium, Cherenkov radiation is emitted. Three photomultipliers look downwards into the water.  The Cherenkov light, produced by muons and electrons (plus pairs and Compton electrons from the much more copious flux of photons), produce light when their energies are above their respective Cherenkov threshold.  Muon energies, to penetrate the tank, must be above 250~MeV (threshold in water $\sim160$ MeV).  The mean electron and photons energies are around 10~MeV (threshold $\sim260$~keV).  The light produced is reflected by a white material and it is then detected by photocathodes. 
\end{textbox}

TA is the largest cosmic-ray detector in the Northern hemisphere \citep{{2008NuPhS.175..221K}}. It is located near the city of Delta, Utah, USA at 1,400m above sea level. It consists of  507 stations, each with two layers of plastic scintillator of 3m$^2$ area deployed in a square grid with a 1.2 km spacing. The total effective area of the SD array is approximately 700 km$^2$. A total of 38 air-fluorescence telescopes are distributed over 3 FD stations grouped at 3 stations. The air-fluorescence telescopes operate on clear, moonless nights and accumulate an average of about 10\% live time \citep{TelescopeArray:2013dpz}. The arrival direction of a cosmic-ray particle as measured by the TA SD is determined from the relative difference in arrival time of the shower front at various SD stations which is measured by time-synchronized GPS modules mounted on each station. The energy estimator of the SD is the particle density measured at a distance of 800 meters from the shower axis, S(800), then converted to an estimate of the primary energy as a function of zenith angle based on a Monte Carlo simulation using the CORSIKA software package \citep{1998cmcc.book.....H}. The  obtained energy is then calibrated to the calorimetric energy measured by the 3 FD stations using a scaling factor of 1/1.27. 

Auger is the largest cosmic-ray detector in the Southern hemisphere \citep{{Abraham2015}}. It is located in the Province of Mendoza, Argentina, at altitudes between 1,340m and 1,610m. The observatory consists of a SD array with 1660 water Cherenkov stations placed in a triangular grid with nearest neighbors separated by 1500m, and a smaller array (stations separated by 750m). The total effective area of the array is approximately 3,000~km$^2$. The SD is made of  Cherenkov water stations and  overlooked by 24 air-fluorescence telescopes grouped at 4 sites. The arrival direction is determined by fitting a spherical model to the arrival times of particles comprising the shower front, described in detail in \cite{2020JInst..15P0021A}. The energy estimate is based on different observable for vertical (events with zenith angles $\theta<60^\circ$) and inclined events ($60^\circ\leq \theta\leq 80^\circ$): the signal at a reference distance of $1000$m from the shower core, $S(1000)$,  for the vertical events,  and  the muon content of the shower with respect to a reference simulated proton shower with energy $E=10^{19}$eV, $N_{19}$, for inclined events. A correction is then applied to take into account the absorption that showers undergo at different zenith angles \cite[described in][]{2020PhRvD.102f2005A}. This method is used to convert $S(1000)$ and $N_{19}$ to the value they would have if the same shower had arrived from a reference zenith angle of $38^\circ$ and $68^\circ$ for vertical and inclined events, respectively. These corrected energy estimators, $S_{38}$ and $N_{68}$, are then calibrated with the calorimetric energy measured by the FD, using a sub-sample of quality hybrid events.

\begin{textbox}[ht]
\section{COSMIC RAY DISTRIBUTION FUNCTION}
It is conventional to introduce a distribution function $f(t,{\bf r},{\bf R})$ to describe the six dimensional phase space (the combination of position space and momentum, or equivalently, rigidity space) density of the relativistic cosmic rays interacting with electromagnetic field. $f$ satisfies the Vlasov equation --- it is constant along a trajectory in phase space. Cosmic rays of rigidity $R$ will conserve their adiabatic invariants when confronted with disturbances with lengthscales larger than their Larmor radii $r_L\sim R/B$ \citep[see, e.g.][and references therein]{2005ppa..book.....K,thorne21}. Resonant waves, with wavelengths $\sim r_L$, scatter the cosmic rays in pitch angle causing them to random walk along the magnetic field. The magnetic field lines also wander causing the cosmic rays to diffuse transversely. Both effects can be captured by introducing a scattering mean free path $\ell(R)$ and a diffusion coefficent, or more generally, a tensor $D\sim\ell c/3$.

In a given inertial frame the scatterers will move with an average velocity $\bf u(t,{\bf r})$, usually that of the background gas. By expanding the distribution function in powers of $u/c$, it can be shown that the lowest order, isotropic part, satisfies the conservation equation
\begin{equation}\label{eq:CDE}
\partial_tf+\nabla\cdot{\bf F}+R^{-2}\partial_R(R^2G_R)=0,
\end{equation}
where the flux of cosmic rays with rigidity $R$, in position space is ${\bf F}={\bf u}f-D\nabla f$ and the flux  in rigidity space is $G_R=-(\nabla\cdot{\bf u})Rf/3$, allowing the cosmic rays to gain momentum adiabatically in a converging flow. The quantity $\bf F$ is not the same as the flux of cosmic rays that an observer would measure at a given momentum. This is given by 
\begin{equation}\label{eq:SJC}
{\bf H}=-{\bf u}R\partial_Rf/3-D\nabla f.
\end{equation}
\end{textbox}

The primary particles can only be studied in statistically large samples, because of the fluctuations in the individual shower development due to changing atmospheric conditions and the fact that the physical processes involved are stochastic. Nevertheless, some systematic behavior can be found as correlated with the energy and type of the primary particle. For example, showers initiated by a high energy primary penetrate deeper in the atmosphere than less-energetic ones. For many EAS of similar primary energy, (which is determined using hybrid events, since the FD enables a quasi-calorimetric measurement of the shower energy), we can define a mean atmospheric depth, $\langle X_{\rm max}\rangle$, and a variance, $\sigma({X_{\rm max}})$, both of which depend on the mass of the primary particle. 

According to the shower superposition principle \citep[see][for a recent account]{2022JPhG...49c5201W}, an EAS initiated by a nucleus with energy $E$ and mass number $A$ can be approximated as the superposition of $A$ proton-initiated showers, each with an energy of $E/A$.  Nucleon-induced showers are $\approx A$ times shallower  than  proton-induced showers by a factor $\sim\lambda_r \ln A$, where $\lambda_r$ is the radiation length in air ($\sim37$ g cm$^{-2}$),
and the variance $\sigma({X_{\rm max}})$ is approximately $\sqrt{A}$ times smaller.   Since the moments of the distributions are used, the mass composition cannot be determined on a event-by-event basis. Together with the moments of the $X_{\rm max}$ distribution, the muon number is a key observable to determine the mass composition of the primary particle because heavy nuclei produce more muons than light nuclei. 

\begin{textbox}[ht]
\section{WAVE SCATTERING AND COSMIC RAY DIFFUSION}
A single resonant MHD wave, with wave vector $k\sim r_L^{-1}$ and small amplitude $\delta B$, will induce a scattering in pitch angle $\sim\delta\phi\sim\delta B/B$ every gyration. If the waves are broadband and randomly phased, then the cosmic rays will diffuse in pitch angle at a rate $\nu=<\delta\phi^2>c/r_L$. The corresponding mean free path is $\ell\sim c/\nu$. Alfv\'en waves are usually supposed to belong to a weak MHD turbulence spectrum created by larger scale fluid motion that cascades down to smaller, resonant scales, typically with ``Kolmogorov'' energy density per $\ln k$, ${\cal U}_{\ln k}\propto k^{-2/3}$. In this case, $D\propto R^{1/3}$. Alternatively, the turbulence may be self-generated by the cosmic rays,especially at shock fronts, as they try to stream through the background gas and cause instabilities to grow to greater amplitudes than those of the background turbulence. The growth rates can be computed perturbatively using ``quasilinear theory''\citep{Jokipii66,2005ppa..book.....K}. However the efficacy of cosmic ray acceleration suggests that the transport is strongly nonlinear. This effect is commonly called Bohm scattering, appropriating a term from thermal plasma physics. Bohm scattering presumes that $\ell\sim r_L$ at all relevant rigidity, from $\sim\,1\,{\rm GV}$ to $\sim10\,{\rm EV}$, when the interaction is strongly nonlinear and there may be no ``uniform'' background field. A natural conjecture is that ${\cal U}_{\ln k}\sim P_{\ln R}(R\sim(8\pi {\cal U}_{\ln k})^{1/2}/k)$ where the associated cosmic ray pressure is $P_{\ln R}\sim4\pi R^4f(R)c/3$.

For Kolmogorov turbulence, $D$ is well approximated by a fitting function taking into account both  the resonant and non-resonant diffusion regimes \citep{2008A&A...479...97G}, $D\approx0.03\left[{\lambda_{\rm Mpc}^2 R_{\rm EV}}/{B_{\rm nG}}\right]^{{1}/{3}}+0.5\left[{R_{\rm EV}}/{(B_{\rm nG}\lambda_{\rm Mpc}^{0.5})}\right]^{2} {\rm Mpc^2 Myr^{-1}}$ where $R_{\rm EV}$ is its rigidity in EV, $B_{\rm nG}$ the field variance in nG and $\lambda_{\rm Mpc}$ its coherence length in Mpc.

In an alternative description of cosmic ray transport, the magnetic field is supposed uniform for long segments, adjoined by ``switchbacks'' \citep{2005ppa..book.....K} where its direction changes abruptly \citep{lemoine23, kempski23}, scattering a broad rigidity spectrum of cosmic rays.  Such structures have been reported in the interplanetary medium. 
\end{textbox}

\subsection{Through the astrophysicist's eyes: ``What is their origin?''}\label{sec:origin}
To the astrophysicist, UHECR present the challenge of identifying the Universe's most  powerful particle accelerators now seen through multi-messengers \citep{1990cup..book.....G,Murase19}. It is difficult to determine the origin of UHECR because magnetic fields filling  the interstellar and the intergalactic medium   deflect charged particles and erase the ``memory'' of their initial directions. Without magnetic fields, UHECR would propagate in straight lines and their arrival directions would point to their sources.  So one crucial question is: {\it ``How high must the rigidity be that we can carry out cosmic ray astronomy?''}. The answer depends  on the strength of the Galactic and extragalactic magnetic fields and their level of turbulence. 

The average strength of the Galactic Magnetic Field, GMF, is $\sim6\,\mu{\rm G}$ near the Sun and increases to $\sim30\,\mu{\rm G}$ in the Galactic center region \citep[][]{2017ARA&A..55..111H}.  It is usually assumed that UHECR with energies above 1 EeV must be extragalactic in origin since their Larmor radius, $r_L\approx1.08 \,\,E_{\rm EeV}/(Z B_{\mu \rm G})\,\,{\rm kpc}\approx R_{\rm EV}/B_{\mu\rm G}\,{\rm kpc}$, is much larger than the size of the Galactic disk. However, our knowledge of the magnetic structure of the Milky Way   in the Galactic halo region is particularly limited, and a large scale turbulent field may be present but invisible, and therefore Galactic models \citep[e.g.,][]{2011ApJ...742..114P, 2016ApJ...821L..24E, 2024PhRvD.110b3016Z} cannot be totally ruled out. The magnetic structure around our local group is even more poorly probed. Hence the energy at which the transition from Galactic-to-extragalactic cosmic-rays is not well established, but some understanding can be reached  from the interpretation of the features observed in the UHECR spectrum (Section~\ref{observations}). It is reasonable to assume that UHECR are extragalactic at the highest energies and as such, the effects of both Galactic and extragalactic media need to be taken into account in modeling their propagation from their source to the Earth.

 \begin{figure*}
\includegraphics[width=5.5in]{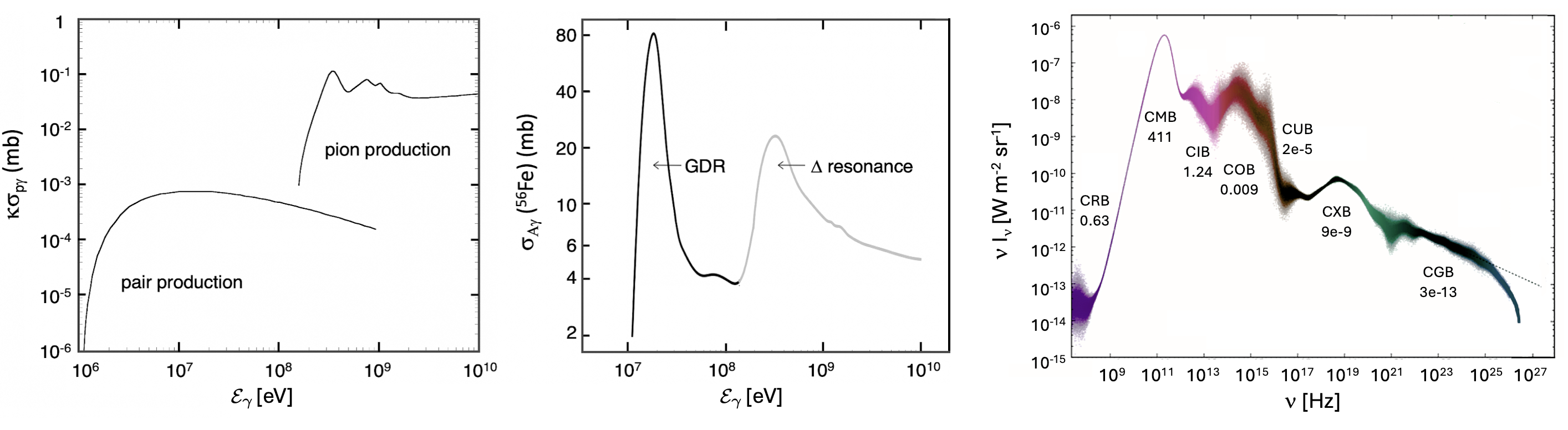}
\caption{{\it Left:} Product of the cross section and the inelasticity $\kappa$ for protons photointeractions (pair production and pion production, respectively) adapted from \citet{2012APh....39...33A}; {\it Middle:} Cross section for GDR and $\Delta$ resonance for iron nuclei, adapted from \citet{2019JCAP...11..007M}; {\it Right:}  Complete cosmic background radiation adapted from \citet{2018ApSpe..72..663H}. The acronyms CRB, CMB, CIB, COB, CUB, CXB, CGB denote the radio, microwave, infrared, optical, ultraviolet, X-ray, and gamma-ray backgrounds, respectively. The numbers indicate the photon number density for each component. The CMB is the dominant photon background with density $\sim411$ photons cm$^{-3}$ and average photon energy $\langle {\cal E}_\gamma \rangle \simeq 6\times10^{-4}$~eV. 
\label{fig:pgammacross}}
\end{figure*}
 
\begin{figure*}
\includegraphics[width=4.7in]{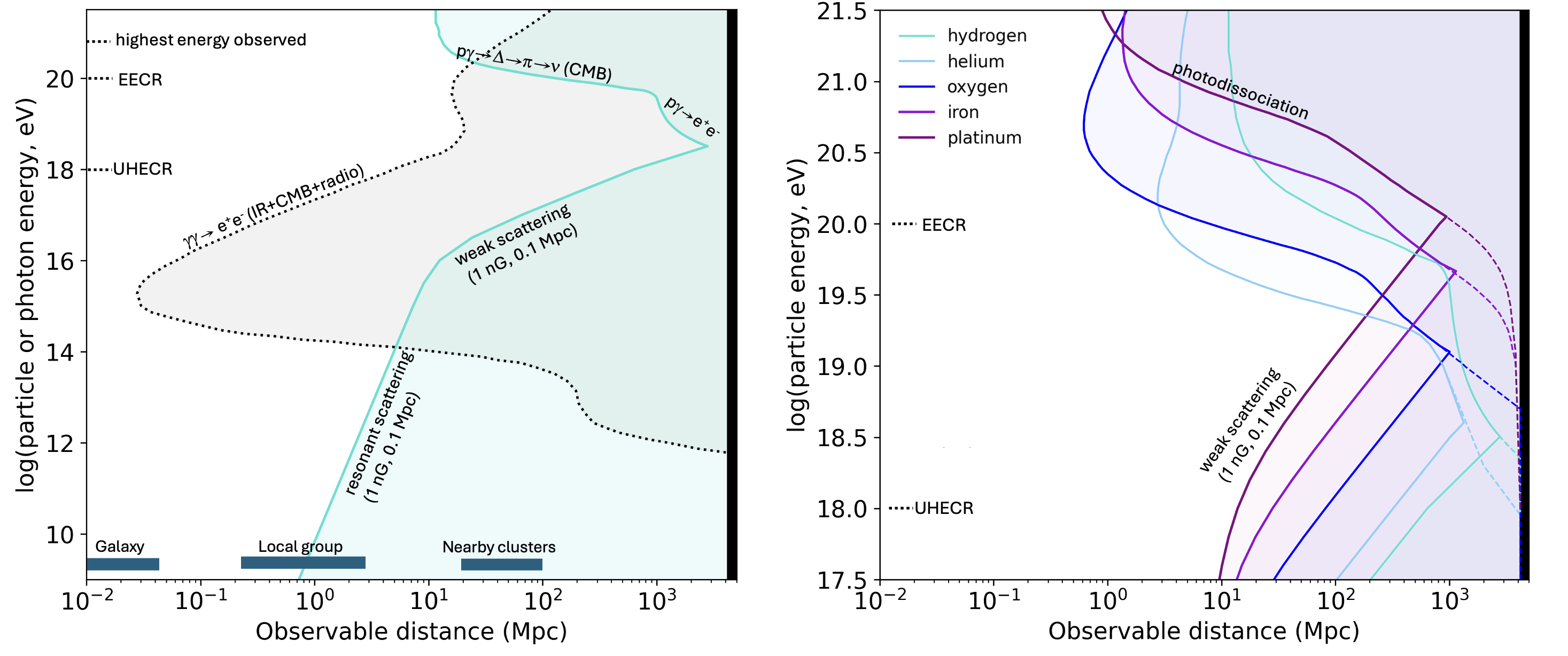}
\includegraphics[width=3.5in]{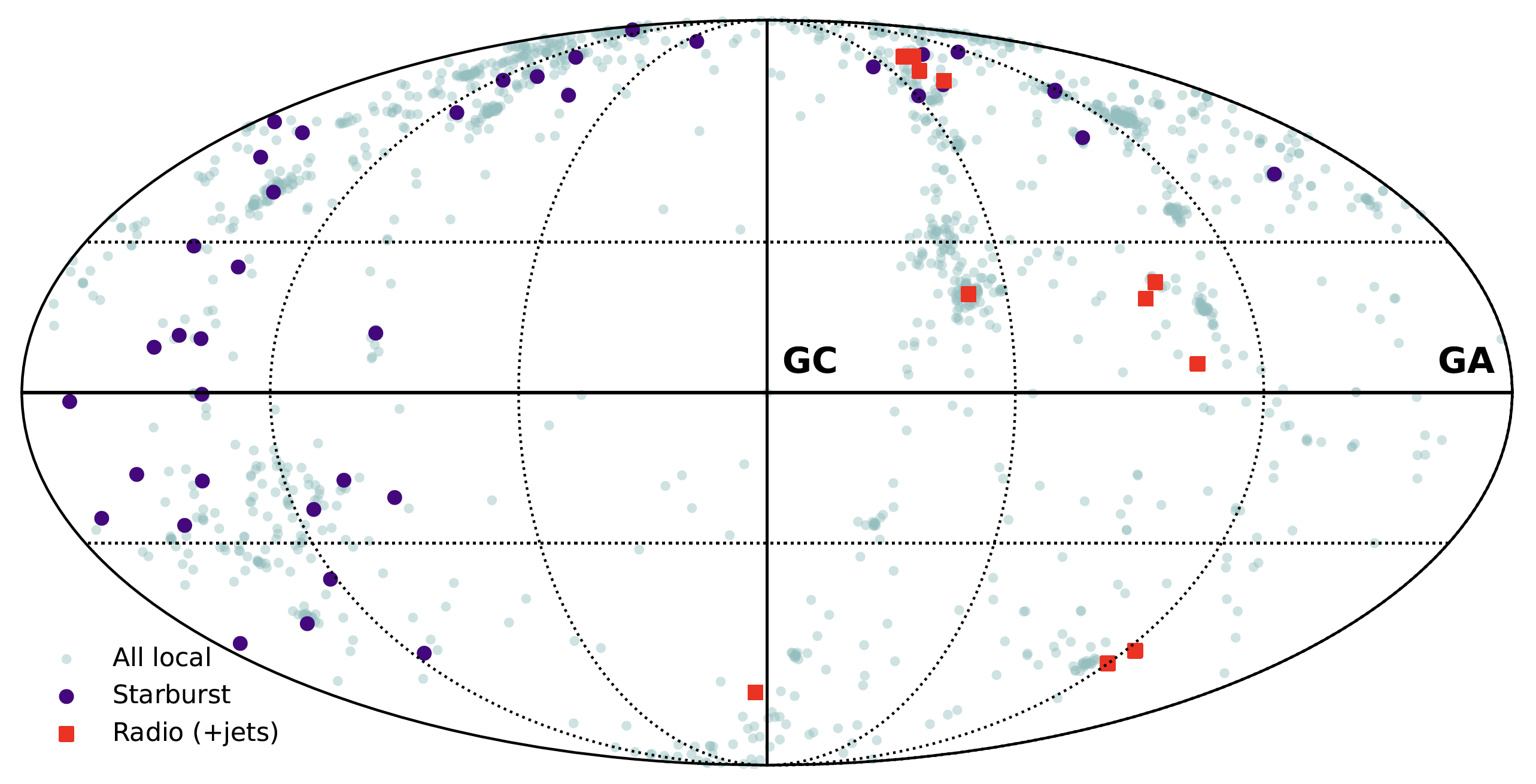}
\caption{{\it Top Left}: Energy evolution of the energy loss length of protons and photons for different energy loss processes in the CMB and IR photon backgrounds at z=0.  {\it Top Right}: Evolution of the attenuation length of iron nuclei as a function of the energy for the different photodisintegration processes and interactions with the CMB and IR/Opt/UV photons at z = 0. At low energy the diffusion distance in the turbulent EGMF is shorter than the loss length (see \citep{2008A&A...479...97G} for details) and UHECR  may be hidden by magnetic horizons. The energy loss lengths for all the  photointeraction processes are taken from \citet{2012APh....39...33A} and \citet{2024arXiv240517409Z}. {\it Bottom}: Sky map in Galactic coordinates of the different types of galaxies within  40 Mpc that could host the sources of UHECR, adapted from \citep{2023ApJ...945...12G}.}
\label{horizon}
\end{figure*}

\begin{textbox}[ht]
\section{COSMIC-RAY HORIZONS}
We can define a parameter similar to an opacity $\tau_{\rm loss}\equiv{t_{\rm prop}}/({\rm min}({t_{\rm loss}},{t_{\rm Hubble}}))$, where $t_{\rm Hubble}$ is the Hubble time, $t_{\rm loss}(E)=-\frac{1}{E}\left(\frac{dE}{dt}\right)$ is the energy loss time and $t_{\rm prop}$ is the propagation time of the UHECR in the magnetic fields (either at the acceleration site, in the source environment or in the extragalactic medium). 

Let us define $t_{\rm loss}$. In the case of hadronic  ($p-p$) interactions,  we have $t_{\rm loss}=(n_p\, c \, \kappa \, \sigma_{pp} ) ^{-1}$ where  $\kappa\equiv dE/E$ is the inelasticity, $n_p$ the proton density and $\sigma_{pp}$ the cross section for hadronic interactions (Fig.~\ref{fig1}).  For photohadronic ($p-\gamma$) interactions, $t_{\rm loss}=(n_{\gamma}\, c \, \kappa \, \sigma_{p\gamma} ) ^{-1}$ (or $t_{\rm loss}=(n_{\gamma}\, c \, \kappa \, \sigma_{N\gamma} ) ^{-1}$ for nuclei) where $n_\gamma$ the photon density and $\sigma_{p\gamma}$ ($\sigma_{N\gamma}$ fpr nuclei) the cross section for hadronic interactions (Fig.~\ref{fig:pgammacross}). 

When $\tau_{\rm loss}\leq 1$ the universe is ``transparent'' to cosmic rays. Therefore the condition $\tau_{\rm loss}= 1$ sets a limit to the distance UHECR can travel, defined as the ``UHECR horizon'' (Fig,~\ref{horizon}). When ${t_{\rm loss}}<{t_{\rm Hubble}}$ and $\tau_{\rm loss}\gtrsim 1$,  there is  efficient production of secondary gamma-ray and neutrinos. Since these neutral particles travel in straight line, they  may provide a smoking gun signal of UHECR production. 
\end{textbox}

The apparent isotropy of UHECR \citep{Fujii:2024sys} suggests that they diffuse in magnetic fields.  The interaction between a distribution of charged cosmic rays ({\bf Sidebar Cosmic Ray Distribution Function}) and the randomelectromagnetic fluctuations is perturbative, assuming $\delta B/B <1$ ({\bf Sidebar Scattering and Diffusion}). However, large-amplitude turbulence with $\delta B/B \gtrsim 1$ can be found near sources and in particular, shock waves, leading to an efficient trapping of cosmic rays and particle acceleration. This is the essence of Fermi acceleration \citep{Fermi49}, one of the earliest and most widely invoked cosmic acceleration mechanisms. As originally conceived, this postulated a population of interstellar clouds, now replaced by MHD waves, moving with speed $\sim u$, off which cosmic rays would ``bounce'' at a rate $\nu_F$ gaining rigidity by an amount $\sim\pm Ru/c$ in each encounter. The cosmic rays random walk in rigidity space at a rate $d \langle R^2\rangle/dt\sim\nu_F(Ru/c)^2$ or $G_R\sim\nu_F(u/c)^2R$ \citep[e.g.][]{2005ppa..book.....K}.

Extra-Galactic Magnetic Field, EGMF, of strength a few microgauss, have been observed near the center of galaxy clusters; a few tenths of a microgauss are reported near their outskirts. Meanwhile, upper limits on the primordial magnetic field in voids have been derived  \citep{2016cmf..book.....K,2024arXiv240616230M}.  The correlation length, $\lambda_c$, provides an average size for the ensemble of magnetic structures that form the turbulence. Commonly assumed values for the field strength are $\sim 1$~nG, and $\sim100$~kpc the for correlation length.  At low energy the diffusion time of UHECR in the EGMF becomes  comparable to the Hubble time and a suppression in the flux of cosmic rays is expected ({\bf Sidebar  Cosmic-Ray Horizon}), leading to the existence of a (rigidity dependent) "magnetic horizon" \citep{2004NuPhS.136..169P, Lemoine05, Berezinsky07}.  Such a flux suppression depends on the source density and cosmological evolution \citep{2008A&A...479...97G}. In reality, the EGMF strength is expected to be correlated with the different structures, clusters, filaments, and voids \citep{Kotera08}. This may change the propagation of the UHECRs in the field with stronger deflections within regions of stronger magnetic field. 

At the highest energies, UHECR astronomers benefit from a phenomenon which also reduces the size of the  observable universe (and hereby the number of  source candidates)  because   UHECR interact  with the extragalactic background photons, mostly the CMB\footnote{The average baryon density in the extragalactic medium is of the order of one proton per cubic meter, which translates into a collision time for {\it p-p} interaction of $\sim2\,10^{13}$~yr making hadronic interactions  negligible.} and lose energy.   This effect  was predicted by \citet{1966PhRvL..16..748G} and \citet{1966JETPL...4...78Z}, two years after the discovery of the CMB (and is therefore known as the GZK effect). This affects proton and nuclei differently. At the highest energies, protons suffer  from pion production while ultra-high energy nuclei  will be gradually stripped off their nucleons (these interactions will be detailed in Section~\ref{MM}). Therefore, any observed high energy cosmic-ray proton can be the fragment of a heavier nuclei starting its journey at larger distances than the strict horizon for protons. This phenomenon does increase the volume for potential sources \citep[e.g.,][]{Metzger11}; it also suggests that observing heavy nuclei at extreme energies from distant sources is less likely. The observational capability for the mass composition above 100 EeV thus carries critical information about the EECR source distance \citep[e.g.,][]{2023ApJ...945...12G}. 

 The energy loss length of protons and nuclei in the extragalactic background light    and their diffusion lengths in a fiducial  turbulent EGMF is shown in Fig.~\ref{horizon}.  At the highest energies, the effect of the fiducial EGMF on EECR is negligible and they suffer mostly from the GZK effect. Their sources need to be within  100 Mpc at 100 EeV and  40 Mpc at 150 EeV, making their identification possible as the number of credible sources is limited (a map of the local universe within 40 Mpc is shown in Fig.~\ref{horizon}). 
 To discover which of these sources are candidates to be the sources of EECR events, one needs to take into account the EECR propagation in the Galactic magnetic field. The large scale ordered component of the Galactic halo field can both increase and decrease the UHECR flux from a given direction through magnetic focusing/lensing \citep{2000JHEP...02..035H,2015arXiv150804530F}. Currently, there are many different GMF models  \citep{2012ApJ...757...14J, 2017A&A...600A..29T,2024ApJ...966..240X,2024ApJ...970...95U,2024arXiv240702148K} so it is difficult to constrain its effects. 

\section{KEY OBSERVATIONAL RESULTS}\label{observations}
\begin{extract}
``Atmosphere! Atmosphere! Do I look like an atmosphere?'' -- Arletty in {\it H\^otel du Nord}, 1938\footnote{Also the year in which Pierre Auger began his detailed studies of the properties of extensive air showers.} 
\end{extract}

As we have seen in the introduction, the spatial and temporal structure of an atmospheric shower is a unique signature of the physical properties (energy and mass) of the primary particle. In this section, we will  review the current status on the  energy spectrum, mass composition and arrival direction of UHECR as derived from the air shower data accumulated during almost two decades by the two current UHECR experiments - TA and Auger. 

\subsection{Energy}\label{spectrum}

\begin{figure}[h]
\includegraphics[width=5.5in]{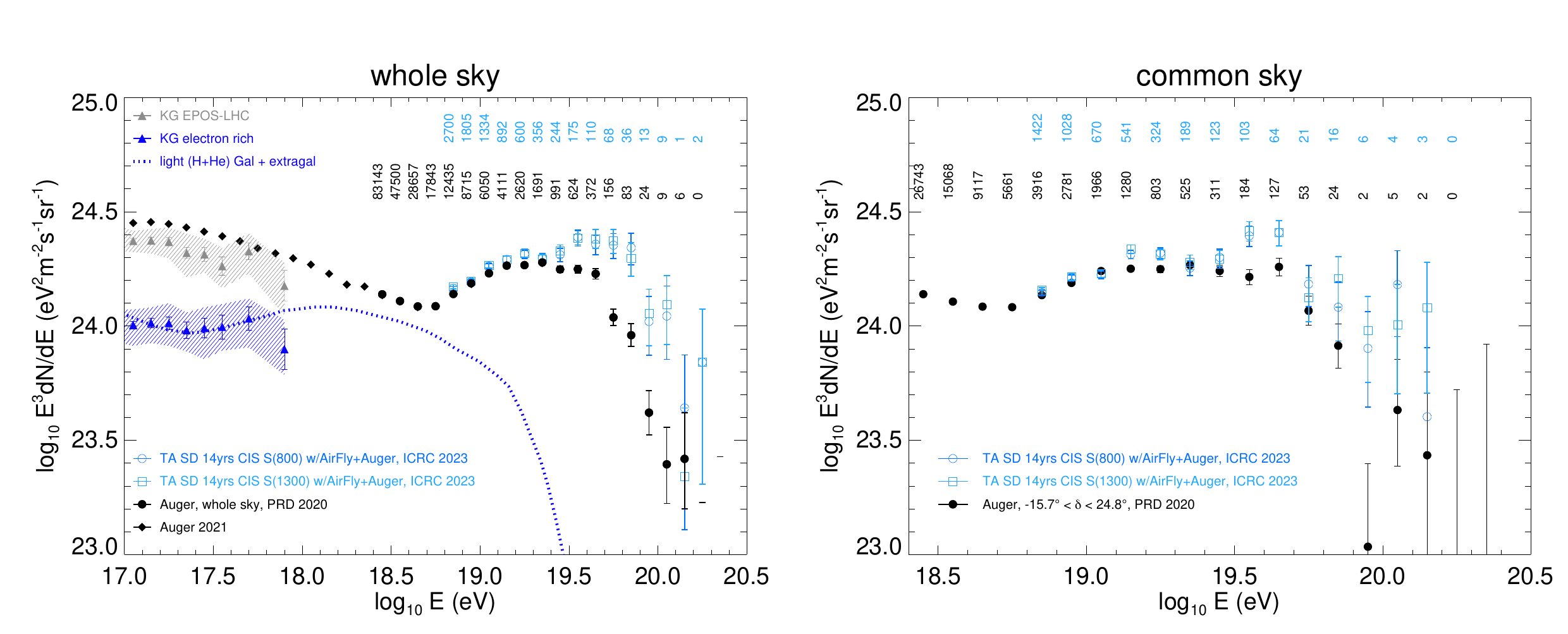}
\caption{{\it Left} Spectrum of UHECR multiplied by $E^3$ observed by KASCADE-Grande, TA and Auger. Overlaid in dotted is the simulated spectra obtained for the proton component which allow to account for the Galactic to extragalactic transition. {\it Right} Spectrum of UHECR multiplied by $E^3$ observed by TA and Auger in the common declination band. The numbers of UHECR in each energy bin is indicated. TA and Auger data are taken from \citep{Bergman23}. KASCADE-Grande data are taken from \citep{KASCADE-Grande:2015bch}. The total UHECR light component across the Galactic-to-extragalactic transition  is shown by the dotted blue line  \citep{2015PhRvD..92b1302G}.}
\label{fig:spectrum}
\end{figure}
The UHECR energy spectrum as measured by TA and Auger is shown in the left panel of Fig.~\ref{fig:spectrum}. It can be described as a power-law, with the flux falling by $\sim$ a thousand   for each decade increase in energy. This points to a non thermal process. A closer inspection of  the spectrum  indicates a hardening at around 5 EeV, with a steepening about a decade higher, around 50 EeV. These two features - called ``ankle'' and ``cutoff'' -  have been confirmed by both experiments. The TA and Auger spectra appear, however, to be different in normalization and shape, even in the common declination band, as can be seen in the right panel of Fig.~\ref{fig:spectrum}. A joint TA-Auger working group have made detailed studies on systematic uncertainties and found that the energy spectra determined by both observatories are consistent in normalization and shape if energy-dependent systematic uncertainties (with an energy-dependent shift $\pm10\%  \times \log (E/10\, {\rm EeV})$ for $E> 10$~EeV) are taken into account \citep{Bergman23}. 

The ankle, a hardening seen in the all-particle spectrum at about 5 EeV, has been the subject of various interpretations over the years. The first successful model able to explain the ankle is the dip model \citep{2006PhRvD..74d3005B}, where the ankle appears as a natural feature of the extragalactic UHECR propagation resulting from  the dip in the pair-production losses of protons on CMB, $ p + \gamma_{\rm CMB} \rightarrow p + e^+ + e^-$, as seen in Fig.~\ref{horizon}. This elegant possibility has been excluded by composition measurements, as the main assumption of this model is that the extragalactic UHECR  are mostly protons with a maximal admixture of 15\% of nuclei \citep{2005A&A...443L..29A}.  In mixed composition models, the ankle is generally considered to mark the transition from Galactic to extragalactic cosmic rays \citep{2002APh....17..125S}. There are hints that this transition may even start at lower energies, around 0.1 EeV, where  an ankle-like feature in the spectrum of the light  (proton and helium) elements (also shown in Fig.~\ref{fig:spectrum}) has been reported by KASCADE-Grande \citep{2013PhRvD..87h1101A}. There is also a  phase shift of the dipole anisotropy at around 0.1 EeV  towards the Galactic anticenter direction  \citep{Fujii:2024sys}. These observations suggest that  a light extragalactic component starts to take over the Galactic component at 0.1 EeV.  

The cutoff marking the end of the spectrum was first reported by HiRes in 2008 \citep{2008PhRvL.100j1101A}, and then confirmed by both  Auger \citep{2009arXiv0906.2189T} and TA \citep{2013ApJ...768L...1A}. This cutoff could be due to the GZK effect (Section~\ref{sec:origin}). It should be noted, however, that the steepening occurs earlier than the GZK prediction for a pure proton composition \citep{Watson24}. Another interpretation of this flux suppression is that it is primarily caused by a limit to the rigidity accelerated at the sources. As for now, the  highest energy observed is 320 EeV \citep{1995ApJ...441..144B},  followed by an event at 244 EeV reported last year \citep{2023Sci...382..903T}.There are $\sim60$ events above 100 EeV \citep{Caccianiga23}.  There is no apparent correlations/clustering with nearby source candidates above 100 EeV, which is interpretable as a heavy composition at the highest energies or a very strong local EGMF.

 \begin{figure*}
\includegraphics[width=5.5in]{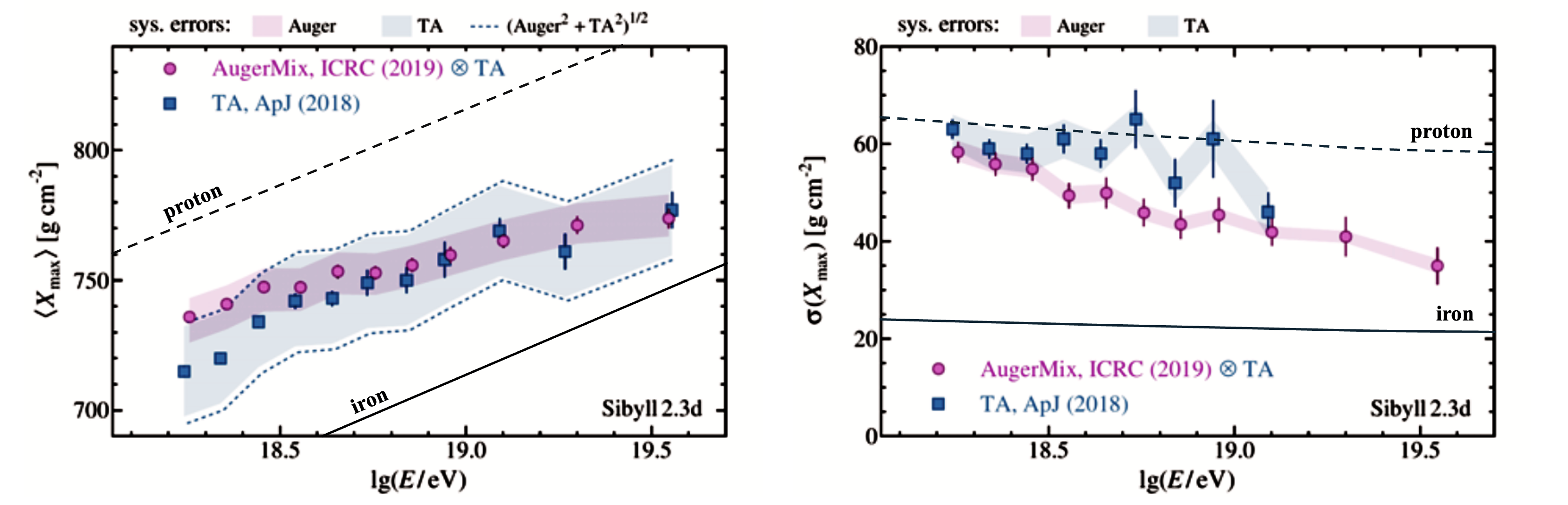}
\caption{The first two moments of the $X_{\rm max}$ distributions. Figure adapted from \citep{2023EPJWC.28302008B} 
shows the energy evolution of the mean (left) and the standard deviation (right) of the $X_{\rm max}$ distributions measured by  TA   (blue squares) and  Auger$\otimes$TA  (red circles) using Sibyll 2.3d. 
 Auger and TA measurements are compatible at the current level of statistics and understanding of systematics \citep{2019EPJWC.21001009Y}. }
\label{fig6}
\end{figure*}

\subsection{Composition}
Over the past decade the evidence against a simplistic, pure, proton UHECR composition has become very strong. Auger revealed a much more nuanced picture where the primary composition is a mixture of protons and heavier nuclei, already at the ankle \citep{2016PhLB..762..288A} and which becomes significantly heavier as the energy increases \citep{2023arXiv231214673S}. This trend towards a heavy composition is also inferred from the  isotropy at the highest energies \citep{2024PhRvL.133d1001A}.  

The evolution of the moments of the $X_{\rm max}$ distribution (see section \ref{detection}) as measured by Auger and compared to the prediction for hydrogen and iron is shown in Figure~\ref{fig6}. 
At 5 EeV, the span in predictions for $\langle X_{\rm max}\rangle$ between the different hadronic interaction models  is $\sim$25 g cm$^{-2}$, 
about one-quarter of the difference between $\langle X_{\rm max}\rangle$ values of  protons and iron nuclei. As consequence, the mass composition of cosmic rays can be referred only with respect to the $X_{\rm max}$ scale predicted by a particular model \citep{2024PhRvD.109j2001A}.

The component below the ankle is required to have a very soft spectrum and a mix of protons and intermediate-mass nuclei \citep{LOFAR,Halim23}. The origin of these "sub-ankle" extragalactic protons is unclear. Photo-disintegration of UHECR nuclei on background photons within or around their sources  was suggested as a viable mechanism for producing these extragalactic protons \citep{Globus2015a, UFA15}. In Galactic models, this light ankle is interpreted as a transition between two light Galactic components \citep{2016ApJ...821L..24E}. The origin of the intermediate-mass component is not well constrained and could be either of Galactic or extragalactic origin.

Above the ankle, the evolution towards a heavier composition 
can be interpreted as the signature of a low maximal energy-per-unit-charge 
of the nuclei accelerated at UHECR sources \citep{2008JCAP...10..033A}. It was quoted as the ``disappointing model'' \citep{2011APh....34..620A} since the low maximum rigidity  would challenge the possibility to perform UHECR astronomy.  

At the highest energies, where there are no measurements of composition-sensitive observable with the fluorescence detectors of Auger and TA, the existence of protons or very heavy nuclei (heavier than iron) cannot be ruled out. A very interesting possibility would be the existence of an additional proton-dominated component at the highest energies which could be detected through secondary neutrinos  (Section~\ref{MM}). The muon puzzle (Section \ref{partphys}) tells us that we need to address the origin of the discrepancies between air shower models and mass measurements  \citep[e.g.][]{Scaria2023} to have a better understanding of the nature of UHECR. What can be said today, on the basis of existing shower models, is that replacing heavy nuclei with protons at the highest energy exacerbates the muon problem.

\subsection{Anisotropy}

The discovery of a large scale modulation in right ascension in the distribution of UHECR with energy above 8 EeV has been a breakthrough in cosmic-ray physics \citep{2017Sci...357.1266P}. 
This anisotropy has now reached a statistical significance of $6.9\sigma$ \citep{2024arXiv240805292T}. It can be interpreted as a dipole moment of amplitude $\sim7\%$ at 10 EeV pointing in a direction $\sim113$ degree away
from the GC, with no statistically significant evidence for a quadrupole moment.

\begin{figure*}[]
\includegraphics[width=5.in]{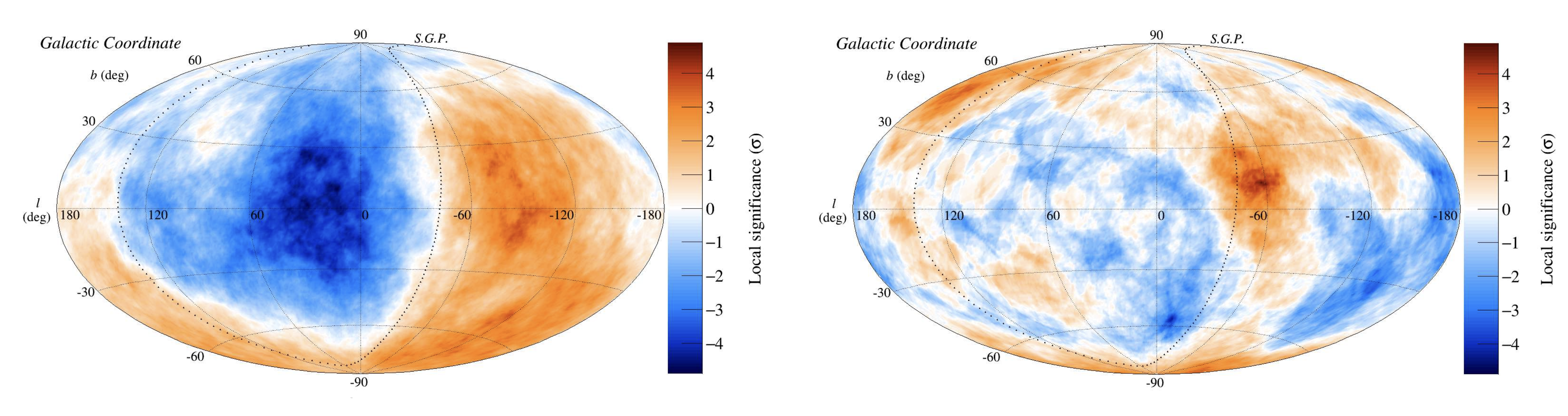}
\caption{{\it Left}: Significance sky-map
of 45$^{o}$ oversamplings of events above the ankle energy. The distribution of UHECR appears to be dipolar with respect to the supergalactic plane, SGP. The dipole is the only significant UHECR anisotropy ($6.9\sigma$). {\it Right}: Significance sky-map of 20$^{o}$ oversamplings above the cutoff energy. The post-trial significance of the intermediate-scale anisotropies are $\lesssim4\sigma$. The sky maps are in in galactic coordinates \citep{2022ICRCFujii}.}
\label{fig:anisotropy}
\end{figure*}

A working group with members from both Auger and TA  have  searched for anisotropies in the UHECR arrival direction data. A full-sky coverage is achieved by combining Auger and TA data \citep[see Fig.~\ref{fig:anisotropy} adapted from][]{2022ICRCFujii}. While the angular resolution is typically better than a degree,  combining the two datasets requires a cross-calibration procedure due to the different systematic uncertainties on energy measurements. Spherical harmonic moments are well-suited for measuring the anisotropy level at different angular scales but the reconstruction of the multipolar moments depends on the energy cross-calibration between the two experiments.
 
Since its discovery, attempts to explain the dipole anisotropy have been made, invoking a large number of UHECR sources following the local inhomogeneities, which require an EGMF below some model-dependent limit in the $\sim\,{\rm nG}$ range \citep{2017ApJ...850L..25G, 2018MNRAS.475.2519H, Globus19, 2021ApJ...913L..13D, 2022A&A...664A.120A,2024ApJ...966...71B}. The amplitude of the dipole is increasing with energy which would be a natural result of the GZK horizon becoming smaller as the energy increases. The direction of the dipole is sensisitive to the GMF model used to reconstruct the UHECR arrival directions on Earth. Since the dipole direction is more than 50 degrees away from the CMB dipole, large deflections in the GMF are needed which indicates that the composition cannot be predominantly light above the ankle. The near-constancy of the direction
of the dipole anisotropy with energy, if confirmed in the future with more statistics, would constrain the evolution of rigidity with energy.

Various hints for higher-energy, smaller-scale anisotropies have been reported \citep{Golup23,Kim23} with $\lesssim4\sigma$ post-trial significance, as well as a  correlation between events and sources when comparing the directions of UHECR with a catalog of starburst galaxies \citep[4.6$\sigma$ post-trial significance when Auger and TA data are combined, see][]{Caccianiga23}. There have also been claimed associations of the UHECR  with prominent nearby galaxies including the powerful starburst galaxy, M82  and the energetic radio source, Centaurus A \citep{aab18}. Surprisingly, no excess has been found from the Virgo cluster which is among the most promising source candidates. This has been dubbed the “Virgo scandal”. It has been shown that this could be the result of a strong demagnification of the sources in the Virgo cluster region, due to the lensing effets of the large scale GMF \citep{2015arXiv150804530F} but this effect is strongly GMF model-dependent \citep{2024A&A...686A.292A}.   

In summary, with the exception of the dipole in the 8-16 EeV energy range, there is no compelling evidence (i.e., with $\geq5\sigma$ post-trial significance) for any intermediate or small-scale anisotropy in the distribution of UHECR or association with individual sources. This stipulation may be relaxed if there is a compelling, multi-messenger, identification (Sec.~\ref{MM}). The problem is compounded by our incomplete understanding of the GMF and EGMF. Note that usually the turbulence is considered to be frozen. However,  UHECR hotspots may ``dance'' due to our changing position in the turbulence \citep{Globus16}.  

\begin{textbox}[!h]
\section{PHOTOHADRONIC INTERACTION}
Photohadronic interactions,  the  interaction of a hadron (proton, $p$, or nucleus, $N$) with a photon are:

{$\bullet\,$\it Bethe-Heitler $e^+e^-$ pair production:}
Protons and nuclei can   undergo the process of pair production, $p(N)+\gamma\rightarrow p(N)+e^++e^-$, as the interacting photon energy reaches $\sim 2m_e$ in the proton (resp. nucleus) rest frame. The energy loss rate of the pair production process for nuclei is, to good approximation, proportional to ${Z^2}/{A}$ (at a given Lorentz factor) meaning that once the energy threshold has been reached, nuclei loose energy more rapidly than protons by producing pairs.

{$\bullet\,$\it Photo-pion/Photo-meson production:}
At higher energy, as the interacting photon energy reaches $\sim m_\pi$ in the proton rest frame, a pion can be created by the interaction of a proton and a photon $p + \gamma \rightarrow p + \pi $. Unlike pair production, this process has a large inelasticity $\kappa\sim0.2$. This process becomes important as the $ p\gamma $ cross section increases rapidly for photons with energies above $ {\cal E}_{\gamma, \rm th} \approx {1.4\times 10^{17}{\rm eV^2}}/{E_p}$. 

{$\bullet\,$\it Giant Dipole Resonance:} 
The process with lowest energy threshold and highest cross-section for photodisintegration is the Giant Dipole Resonance, GDR. The GDR cross section is roughly proportional the mass number $A$ of the nucleus, $\sigma_{GDR}=1.45 A\,10^{-27}$ cm$^2$. Most nuclei have an energy threshold $\Delta\epsilon_{\rm GDR}=8$~MeV (except He with a threshold  $\sim20$~MeV and $^9{\rm Be}\sim1$~MeV). The nucleus responds to an electromagnetic excitation by emitting one or a few nucleons. 

{$\bullet\,$\it Higher energy photodisintegration processes:} 
The Quasi-Deuteron process takes place for photon energy above $\simeq 30$~MeV in the nucleus rest frame and can be seen as the emission of a virtual pion inside the nucleus which results in the emission of a pair of nucleons. The residual nucleus is in general excited and may emit more nucleons before reaching its ground state. Above  $\epsilon\simeq m_\pi$ in the nucleus rest frame,  a pion can be created by interaction of the incoming photon through a process called a Baryonic Resonance. The pions created can either escape or be absorbed in the nucleus. This process usually triggers the emission of several nucleons.
\end{textbox}

\section{MULTI-MESSENGER CONSTRAINTS}\label{MM}
\begin{extract}
``The universe is transparent in the direction of the future.''— Hubert Reeves
\end{extract}

\subsection{General considerations}
As the universe expands and cools, it becomes more and more transparent to UHECR and as a consequence their GZK horizon grows. Although the UHECR we observe directly are local (the GZK horizon is $\sim$100~Mpc at $z=0$), UHECR may still be accelerated at cosmological redshifts and their sources may be observable indirectly through secondary neutrinos and gamma-rays. These can be produced either in intergalactic space, as a result of propagation in the extragalactic background photons (``cosmogenic'') as by-products of the GZK effect or within the sources (``astrophysical'') through either hadronic or photohadronic interactions.

\subsection{Cosmogenic constraints}
UHECR contribute to the extragalactic gamma-ray  \citep{Ackermann15} and neutrino \citep{2022JHEAp..36...55A} backgrounds through GeV-TeV photons  produced mostly through the pair production process \citep{1973Ap&SS..20...47S,1973Natur.241..109S}, and PeV--EeV neutrinos produced through the decay of charged pions \citep{1969PhLB...28..423B,waxman13}. The proton content of UHECR is a crucial quantity that affects the fluxes of cosmogenic photons and neutrinos \citep{2011A&A...535A..66D}, respectively. Heavy nuclei photo-disintegrate before producing pions and thus protons are expected to produce significantly more cosmogenic neutrinos than heavier nuclei ({\bf Sidebar Photohadronic Interactions}). 

\begin{figure*}[]
\includegraphics[width=5.7in]{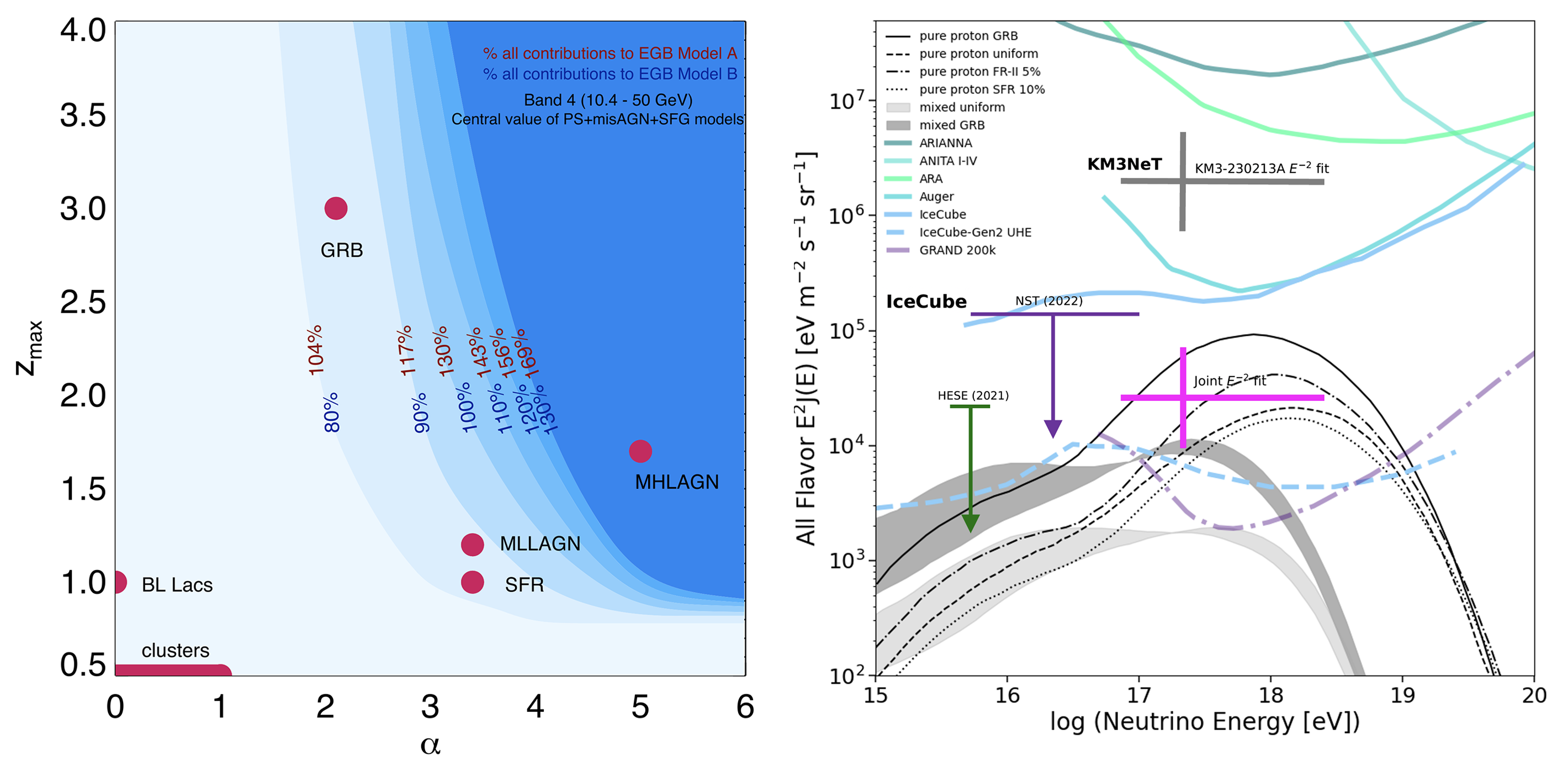}
\caption{{\it Left panel.} Fermi-LAT constraints on UHECR source evolution in the case of a mixed-composition UHECR source model with a soft proton component  \citep{globus17b}. The different shades of blue show the percentage of the sum of point sources, UHECR, misaligned AGNs, and star forming galaxies (UHECR+PS+misAGN +SFG) to the extragalactic gamma-ray background (EGB, models A and B of \citet{Ackermann15}) in the 10.4–50 GeV energy band. The source evolution is characterized by ($z_{\rm max}$, $\alpha$) where $z_{\rm max}$ is the maximum redshift up to which sources experience a cosmological evolution in $(1 + z)^\alpha$.  GRB: gamma-ray bursts. SFR: star formation rate. MLLAGN: medium-low-luminosity AGNs. MHLAGN: medium-high-luminosity AGNs. (High-luminosity AGNs with $\alpha\sim7$, $z_{\rm max}\sim1.7$ do not appear on the figure). 
{\it Right panel}.  Predicted all-flavour cosmogenic neutrino background for various UHECR source evolution  and assuming pure protons and mixed composition  \citep{globus17b} against the expected differential 90\% C.L. sensitivities for a variety of experiments to an all-flavor diffuse neutrino flux, computed in decade-wide energy bins and assuming a ten-year integration. The solid lines show experimental upper limits at higher energies from  Auger \citep{Aab19}, ARA \citep{ARA20}, ARIANNA \citep{ARIANNA20}, ANITA I-IV \citep{ANITA19}, and IceCube \citep{IceCube17}. The dotted dashed line show the  GRAND sensitivity with 200,000 stations \citep{GRAND20}. Sensitivity to UHE neutrinos from IceCube-Gen2 \citep[dashed line, Fig. 15 of][]{2022JHEAp..36...55A} is computed using radio and optical for 10 years. The  KM3NeT-only measurement and the joint flux including as well IceCube-EHE and Auger non-observations, in the central
90\% neutrino energy range associated with KM3-230213A, are shown with the grey and pink crosses, respectively \citep{km3net}.  The violet and green arrows denotes the 68\% CL contours of the IceCube single-power-law fits NST \citep{icecube22} and HESE \citep{icecube21}, respectively.  } 
\label{fig:MMA}
\end{figure*}

A  pure-proton composition  for UHECR (also called ``dip model'', see Section~\ref{spectrum})  is excluded by  Fermi measurements of the diffuse gamma-ray flux  \citep{2016ApJ...822...56G}, as well as by the Icecube upper limits on the neutrino background \citep{Heinze16}. As protons are the main contributors to cosmogenic neutrinos, existing limits on, or future detection of, the high energy neutrino background are very important for constraining powerful proton accelerators, even if they are a small contributor to the whole UHECR spectrum. As can be seen in the right panel of Fig.~\ref{fig:MMA}, pure-proton accelerators evolving like high-luminosity radio galaxies are challenged by the current IceCube limits; but if they contribute only 5\% of the total flux, they should be detectable with the future IceCube-Gen2 or GRAND200k. For strong source evolution, $\propto(1+z)^4$, the current limits of IceCube and Auger constrain the proton fraction to $\lesssim$20\%,  \citep{Aab19,2024JCAP...02..022E}. 

In the case of a mixed composition, low-$E_{\rm max}$ scenarios with a soft proton spectrum and a hard spectrum for the nuclei account for the evolution of the composition seen by Auger. The soft  proton component responsible for the light ankle, represented by the dotted blue line in Fig.~\ref{fig:spectrum}),  is the dominant contributor to the cosmogenic gamma-ray flux \citep{globus17b}.  The left panel of Figure~\ref{fig:MMA} shows the allowed parameter space of different evolutionary scenarios for such a mixed composition model.  The plot is shown for the 10-50 GeV band where the UHECR contribution to the extragalactic gamma-ray background is the largest. The Fermi observations rule out source models with large cosmological evolution such as high luminosity radio galaxies. The corresponding cosmogenic neutrino fluxes, for GRB and non-evolving sources, are shown in the right panel. For a mixed composition, neutrino fluxes are well below the current IceCube limits, and only within the reach of the most sensitive planned future observatories. 

\subsection{Astrophysical constraints}
 Diffuse fluxes of astrophysical neutrinos are also expected from interactions of UHECR in and around their sources \citep{Muzio22}.  Neutrino flux estimates  for  different class of source have been made: clusters of galaxies \citep{Kotera09, Fang18},  high luminosity Gamma-Ray Bursts, GRB \citep{Globus2015a}, low luminosity GRB \citep{Boncioli19, Biehl18a}, binary neutron star mergers \citep{Gottlieb21}, Tidal Disruption Events, TDE \citep{Biehl18b}, newborn pulsars \citep{Fang14}, millisecond magnetars \citep{Fang17}, engine-driven supernovae \citep{Zhang19}, starbursts galaxies \citep{2023PhRvD.107h3009C}, Active Galactic Nuclei, AGN \citep{Rodrigues21,Mbarek23}. Some of these sources are radiation-dominated and thus naturally account for the light ankle (Section~\ref{spectrum}) as a result of the photodissociation of nuclei inside or around the source \citep{Globus2015a, UFA15, Fang18, Muzio23} which has a contribution to the gamma-ray background \citep{globus17b}. Some of these astrophysical neutrino flux estimates  flirt with the current sensitivities of IceCube and Auger. However, it should be kept in mind that all these estimates are quite uncertain, given our poor understanding of the gas and photon fields around the sources and the confinement time of the cosmic rays.  

There have been tentative identifications of $\sim$PeV neutrinos with individual astrophysical sources and classes of sources, including Seyfert galaxies, notably NGC~1068 \citep{ngc1068},  AGN jets, notably TXS 0506+056 \citep{TXS}, TDE \citep{nutde}. If these are correct, then they provide strong circumstantial evidence of processes which may be linked to the production of UHECR at much higher energy. However, none of these identifications is secure and they all raise serious questions. Suppose we have a putative source at distance $r=r_{\rm source}$ and the horizon for this messenger is at a distance $\sim r_{\rm horizon}$, ignoring cosmological corrections. If the sources are steady with constant density, the probability per unit $r$, within a given solid angle defined by the angular resolution, typically degrees, of obtaining detecting a gamma-ray or neutrino is uniform. For example, for every neutrino we associate with NGC~1068, at a distance of $r_{\rm source}\sim14\,{\rm Mpc}$ there should be $\sim300$ more out to $r_{\rm horizon}$ of order a Hubble radius from other sources within the same solid angle. Cosmological evolution exacerbates this problem and would make a strong identifications very surprising. The problem remains if the identification is with a catalog of local sources and is even less convincing if this catalog is considered, {\it a posteriori}, after the observations are made. 

Another major challenge is the fact that the $\sim$ TeV-PeV neutrinos typically come with 80-30\% probability of being atmospheric, which further reduces the significance of possible associations. The best way to make secure source identifications are through temporal coincidences \citep[e.g.][]{waxman13,Yannis21}. This is especially important for transient sources like GRB. Alternatively, identifications are easier for gamma-rays with energy $\gtrsim100\,{\rm TeV}$, where $r_{\rm horizon}\lesssim10\,{\rm Mpc}$. An exciting, recent development is the report by KM3NeT of the detection of a $\sim120\,{\rm PeV}$ muon perhaps associated with a neutrino of energy $\sim 220\,{\rm PeV}$ \citep{km3net}.  If more such events are validated, then they should be highly relevant to identifying the sources of UHECR.

\section{CHALLENGES, SOURCES AND MECHANISMS}
\begin{extract}
``Have you got some reason you want my atoms scattered all over space, boy?'' -- McCoy to Data, in {\it Star Trek} 
\end{extract}
Who is the real culprit flinging UHECR at us, before they are scattered all over space by cosmic magnetic fields? After half a century, many, very different source models are still being considered for the acceleration of UHECR \citep{Coleman2023}. 
Almost all of them assume that the sources are extragalactic; some place them in the intergalactic medium; others associate them with galaxies.   Some sources are quasi-steady on a cosmological timescale; others are transient over a human lifetime. We can summarize the foregoing discussion by stating seven challenges to models that purport to account for the provenance of UHECR. We will then review the different source candidates and  mechanisms with respect to those challenges, as summarized in Table~\ref{table1}.

\noindent$\bullet$ {\bf Rigidity Challenge.} The simplest and most widely adopted interpretation of the highest energy particles, the EECR, with energy up to $\sim300\,{\rm EeV}$, is that they are heavy nuclei, like iron, with rigidity up to $R_{\rm max}\sim10\,{\rm EV}$. The rigidity can be thought of as an electric potential difference that individual particles traverse as they are accelerated. (It is the electric field, that changes the energy of a charged particle.) It turns out that, in essentially all of the acceleration schemes we consider, the maximum rigidity that can be accelerated, $R_{\rm EM}$ can be related to electromagnetic power $L_{\rm EM}$ flowing into the accelerator using an effective resistance $Q_{\rm eff}$\footnote{This quantity is model-dependent and might provide a good target for simulations of different acceleration sites.}, (using $Q$ instead of $R$ to avoid confusion with the rigidity), through
\begin{equation}
R_{\rm max}\sim(L_{\rm EM}Q_{\rm eff})^{1/2},
\label{eq1}
\end{equation}
where the rigidity is measured in volts, the resistance in ohms and the power in watts \citep{blandford00}. This expression is essentially equivalent to the helpful and influential Hillas criterion \citep{1984ARA&A..22..425H}, but with the advantage that the limit depends on one variable $L_{\rm EM}$, instead of two, namely the magnetic field, $B$ and the size, $r$\footnote{The Hillas criterion is that the Larmor radius be smaller than the size, $r$, of the accelerator. This implies that the maximum rigidity, $R_{\rm max}\sim Br$. The square of this involves the magnetic force which is $\propto\langle B^2\rangle r^2$.}. Satisfying this acceleration condition is necessary, though insufficient to meet the rigidity challenge. There may also be significant losses happening both during the electromagnetic acceleration and during escape from the vicinity of the source. This may result in a significantly lower value of $R_{\rm max}$.

\noindent$\bullet$ {\bf Luminosity Challenge.} A successful model must also account for the inferred spectral luminosity density, ${\cal L}_{\rm UHECR}\sim{\cal U}_{\rm UHECR}/t_{\rm loss}$, where ${\cal U}_{\rm UHECR}$ is the cosmic ray energy density measured at Earth with $t_{\rm loss}$ equal to the shorter of the radiation loss time and the cosmological expansion time (Fig.~\ref{horizon}). For iron/proton UHECR, ${\cal L}_{\rm UHECR}\sim3/9\times10^{44}\,{\rm erg\,}{\rm Mpc}^{-3} {\rm yr}^{-1}$; for iron EECR, with $E\sim100\,{\rm EeV}$, ${\cal L}_{\rm EECR}\sim10^{43}\,{\rm erg\,}{\rm Mpc}^{-3}{\rm yr}^{-1}$.  
The luminosity in UHECR per steady source should be consistent with the electromagnetic luminosity responsible for the particle acceleration:
$L_{\rm UHECR}\sim10^{43}n_{-6}$~erg s$^{-1}$ where $n\equiv10^{-6}n_{-6}$~Mpc$^{-3}$ is the source density; or equivalently for bursting sources, the energy released in UHECR should satisfy $E_{\rm UHECR}\sim10^{50}\dot{n}_{-6}$~erg where $\dot{n}\equiv10^{-6}\dot{n}_{-6}$~Mpc$^{-3}$~yr$^{-1}$ is the source rate.

\noindent$\bullet$ {\bf Spectral Challenge.} The observed UHECR spectrum exhibits a light ankle, an ankle and a steepening at high energy. These appear to be accompanied by changes of composition as discussed in Sec.~\ref{spectrum}. If there is one type of source, a successful model should account for a very hard source spectrum for the composed nuclei (harder than $E^{-1}$ below $E_{\rm max}$) with a rigidity-dependent cutoff; a much softer source spectrum for the protons to account for the light ankle.

\noindent$\bullet$ {\bf Injection Challenge.} Essentially all the mechanisms under consideration for accelerating UHECR posit a supply of lower rigidity particles. In some cases, this is naturally part of the input; in others, the injection is local, perhaps from the underlying thermal plasma. The rate of injection needs to match the inferred production of UHECR.

\noindent$\bullet$ {\bf Composition Challenge.} The tentative inference that the EECR are heavy requires a separate explanation in terms of the injection of particles into the acceleration process. Alternatively, it may arise naturally as a consequence of adding energy loss processes. If there is a large population of $\gtrsim300\,{\rm EeV}$ protons, then their sources must be within our local group, seriously impacting the other challenges. 

\noindent$\bullet$ {\bf Anisotropy Challenge.} The distribution of observed sources and the strength of the intergalactic and the Galactic magnetic turbulence must reproduce the observed $\sim7$~percent amplitude of the dipole anisotropy for cosmic rays with $E\gtrsim8\,{\rm EeV}$. Since the effect of the GMF is so model-dependent, it is hard to use the direction of the dipole as a strong constraint at the moment.

\noindent$\bullet$ {\bf Background Challenge.} Although we can only see the highest rigidity cosmic rays from nearby, there could be a multi-messenger, cosmological background --- gamma-rays, neutrinos or even gravitational waves, associated with their sources. Upper limits and measurements of these backgrounds constrain source models, especially when large evolution is expected. 

\begin{table}
\begin{center}
\begin{tabular}{||c||c|c|c|c|c|c|c||}
\hline\hline
{\tiny Models}&{\tiny Rigidity}&{\tiny Luminosity}&{\tiny Spectral}&{\tiny Injection}&{\tiny Composition}&{\tiny Anisotropy}&{\tiny Background}\\
\hline\hline
{\tiny Top-down}&{\cellcolor{green}4.2}&{\cellcolor{green}4.2}&{\cellcolor{red}4.2}&{\cellcolor{orange}4.2}&{\cellcolor{red}4.2}&{\cellcolor{red}4.2}&{\cellcolor{red}4.2}\\
\hline
{\tiny Galaxy Clusters}&{\cellcolor{orange}4.3.2}&{\cellcolor{green}4.3.2}&{\cellcolor{green}4.3.2, 4.3.3}&{\cellcolor{green}4.3.2, 4.3.5}&{\cellcolor{green}4.3.2}&{\cellcolor{orange}2.3}&{\cellcolor{green}4.3.2}\\
\hline
{\tiny Galactic Winds}&{\cellcolor{orange}4.2}&{\cellcolor{orange}4.2}&{\cellcolor{orange}4.0}&{\cellcolor{orange}4.2}&{\cellcolor{green}4.2}&{\cellcolor{orange}4.2}&{\cellcolor{orange}4.2}\\
\hline
{\tiny Magnetars}&{\cellcolor{green}4.4.2}&{\cellcolor{green}4.2}&{\cellcolor{orange}4.0}&{\cellcolor{orange}4.2}&{\cellcolor{orange}4.2}&{\cellcolor{orange}2.3}&{\cellcolor{orange}4.4.2}\\
\hline
{\tiny GRB Jets}&{\cellcolor{green}4.4.3}&{\cellcolor{red}4.2}&{\cellcolor{orange}4.4.3}&{\cellcolor{orange}4.4.3}&{\cellcolor{green}4.2}&{\cellcolor{orange}4.2}&{\cellcolor{green}4.4.3}\\
\hline
{\tiny AGN Jets}&{\cellcolor{green}4.4.3}&{\cellcolor{green}4.4.3}&{\cellcolor{orange}4.4.3}&{\cellcolor{green}4.4.3}&{\cellcolor{green}4.2}&{\cellcolor{orange}4.2}&{\cellcolor{orange}4.4.3}\\
\hline
{\tiny TDE Jets}&{\cellcolor{green}4.2}&{\cellcolor{orange}4.2}&{\cellcolor{orange}4.4.3}&{\cellcolor{orange}4.4.3}&{\cellcolor{green}4.4.3}&{\cellcolor{orange}2.3}&{\cellcolor{orange}4.4.3}\\
\hline\hline
\end{tabular}  
\end{center}\caption{Seven models confront seven challenges, as discussed in the sections indicated by the numbers in the table, rated as: general consistency with observation so far (blue), opportunities for observational and theoretical investigations to resolve differing research findings (grey), ruled out by existing observations (red).}
\label{table1}
\end{table}
\subsection{Seven Challenges}

\subsection{Top-down Cosmological Models}
``Top–down'' is a generic descriptor of models in which the  UHECR primaries are produced as products of hypothetical particles with mass greater than the most energetic EECR with energy  $\sim0.3\,{\rm ZeV}$. As these particles were presumably created in the early universe, at least one stage in their decay chains must have been cosmologically long. A rich variety of possibilities has been entertained, including inflatons ($\sim10-100\,{\rm ZeV}$), topological defects \citep{protheroe96}, Super Heavy Dark Matter particles, to cite a few \citep[see][for details]{2000PhR...327..109B}. If these particles had a cosmological origin they would presumably decay over a cosmological timescale. A fraction $\sim10^{-9}$ of the dark matter energy density would suffice to account for the UHECR luminosity density. There are three serious, generic problems with this otherwise exciting possibility. UHECR should be mostly photons which is explicitly excluded by Auger data \citep{2009APh....31..399P}. Secondly, it is hard to avoid overproducing the measured fraction of the gamma-ray background not accounted for by astrophysical source as measured by Fermi Gamma-ray Space Telescope \citep{ackermann18}. Thirdly, heavy nuclei are not natural end products of these decay chains. As a result, top-down models are currently disfavoured \citep{2008arXiv0810.3017K}.

\subsection{Continuous Acceleration at Non-Relativistic Shock Waves}
\subsubsection{General considerations}
Observations and simulations inform us that the Circum- and Inter-Galactic Medium, CGM and IGM, is very active. Cosmological perturbations became nonlinear and create large amplitude waves which can steepen and form shocks. Gas falls onto galaxies, their groups, filaments and clusters. Stellar processes within galaxies produce outflows which coexist with these inflows. Galactic nuclei are responsible for faster outflows in the form of giant winds and jets which can reach beyond galaxies' virial radii and fill much of the IGM with intermediate-energy cosmic rays and magnetic field. The power/luminosity density needed to account for the UHECR  is only $\sim10^{-6}$ that associated with stars in galaxies. 

The IGM is mostly ionized but highly inhomogeneous \citep[e.g.][]{McQuinn2016}. The mean density is $\sim2\times10^{-7}\,{\rm cm}^{-3}$. Measured temperatures range from $\sim10^5\,{\rm K}$ to $\sim10^7\,{\rm K}$. The sound speeds are then in the range $\sim30-300\,{\rm km\,s}^{-1}$.  Equipartition magnetic field strengths ---  surely upper limits ---  range from $\lesssim1\,{\rm nG}$ in cosmic voids to $\sim30\,{\rm nG}$ in the general IGM. The corresponding Larmor radii for an EECR with $R\sim10\,{\rm EV}$ are $r_L\gtrsim10\,{\rm Mpc}$ in voids, $\sim300\,{\rm kpc}$ in the general IGM. Traditional Fermi acceleration can be ignored.  

However, just as happened with the ISM, attention has turned to shock fronts as localized sites of much more rapid acceleration. The theory of Diffusive Shock Acceleration, DSA ({\bf Sidebar Diffusive Shock Acceleration}) \citep[e.g.][and references therein]{1983RPPh...46..973D,1987PhR...154....1B} has been successfully applied to shocks in the solar wind and those associated with supernova explosions and hot stars. Two intergalactic possibilities for DSA acceleration of UHECR are cluster and filament accretion shocks \citep{Norman95, Kang96, Ryu03, Malkov11, Kim19, Blandford23, Simeon23, simeon25} and galactic wind termination shocks \citep{Jokipii1987, anchordoqui18, Romero2018} as these inevitably process the lower energy cosmic rays escaping from their host galaxies. 

\subsubsection{Cluster accretion shocks}
The closest rich cluster of galaxies is the Virgo cluster, containing several hundred galaxies located at a distance $\sim17\,{\rm Mpc}$. It is therefore near enough to be a UHECR source and may be the dominant contribution to the dipole anisotropy. We use it as an example. More distant clusters can contribute out to $\sim100\,{\rm Mpc}$, depending upon the composition and the level of IGM turbulence (Fig.~\ref{horizon}). For Virgo, simulations suggest a shock radius $r_{\rm shock}\sim2\,{\rm Mpc}$ a preshock density $\sim10^{-29}\,{\rm g\,cm}^{-3}$ and an infall speed $\sim1000\,{\rm km\,s}^{-1}$. The power crossing this shock is $\sim2\times10^{45}\,{\rm erg\,s}^{-1}$. If we adopt a rich cluster density of $\sim10^{-5}\,{\rm Mpc}^{-3}$ \citep[e.g.][]{schneider15}, then the luminosity density associated with the cluster inflows is $\sim3000\,{\cal L}_{\rm UHECR}$. The luminosity challenge is easily met. 

\begin{textbox}[ht]
\section{DIFFUSIVE SHOCK ACCELERATION}
This important particle acceleration mechanism relies on tapping the relative kinetic energy of the plasma on either side of a collisionless shock front. It is observed within the heliosphere and around supernova remnants. Larger and more energetic shock fronts must be found around galaxies and clusters of galaxies. If we regard such shock fronts as adiabatic, planar, and strong, the speed of the downstream gas away from the shock is a quarter of the speed of the upstream gas approaching the shock, $u$. Cosmic rays, traveling with speed $\sim c$, are supposed to be scattered by magnetic inhomogeneities on either side of the shock front moving with the gas. This implies that, on average, they cross the shock front $O(c/u)$ times. As they gain energy systematically by a fractional amount $O(u/c)$ every shock crossing, a typical cosmic ray gains energy by an amount $O(1)$. A kinetic calculation using Eq.~(\ref{eq:CDE}) and imposing Eq.~(\ref{eq:SJC}) as a junction condition at the shock front, leads to the conclusion that incident cosmic rays with rigidity $R_{\rm in}$, will be transmitted downstream as a power law distribution of rigidity $dN/d\ln R\propto R^{-1}$ for $R>R_{\rm in}$. This, in turn, implies that each interval of $\ln R$ carries equal amounts of rigidity/energy for $R>R_{\rm in}$. 

This simple, test particle description has been elaborated by: including the cosmic rays in the shock jump conditions, allowing the cosmic rays to create the turbulence that scatters them, adding acceleration and loss processes and including the shock width, $r_{\rm shock}$ \citep{Eichler84}. Assuming diffusive transport, $r_{\rm shock}$ is also roughly  the thickness of the shock pre-cursor and can be equated to $\sim(c/u)$ mean free paths which is $\sim(c/u)(R_{\rm max}/B(R_{\rm max}))$ assuming the magnetic field is mostly resonant. If we now equate the electromagnetic power $L_{\rm EM}$ to a few times $B(R_{\rm max})^2ur_{\rm shock}^2/4\pi$, we recover Eq.~(\ref{eq1}) with $Q_{\rm eff}\sim30(u/c){\rm ohm}$. Particle in Cell simulations go beyond kinetic theory by tracing the orbits of individual charged particles in the electromagnetic field, continuously summing their contributions to the charge density and current which source the electromagnetic field \citep[e.g.][]{2011ApJ...726...75S,lemoine23}. However, they have not demonstrated the high efficiency needed to meet the rigidity and luminosity challenges especially when starting with thermal particles \citep[][for a review]{Marcowith2020}. 
\end{textbox}

There are two key differences that cluster shocks exhibit from simple, planar shocks accelerating test particles. Firstly, they are outwardly curved. This facilitates the escape upstream of the highest rigidity cosmic rays with Larmor radii $r_L\sim r_{\rm shock}(u/c)$. Up to half of the highest rigidity cosmic rays eventually escape upstream and can be observed. The cosmic rays transmitted downstream, together with the accompanying hot gas and magnetic turbulence, will sink subsonically into the cluster and can only be observed indirectly. Secondly, in order to meet the rigidity challenge, the cosmic-ray and magnetic energy densities close to the shock must be orders of magnitude larger than those far upstream. In our example, $Q_{\rm eff}\sim 0.1\,{\rm ohm}$ and $R_{\rm max}\sim5-10\,{\rm EV}$, with downstream pressure $\sim7\times10^{-14}\,{\rm dyne\,cm}^{-2}$, equipartition equivalent magnetic field $\sim1\,\mu{\rm G}$ and the associated Larmor radius is $\sim6\,{\rm kpc}$, much smaller than in the general IGM.  Richer clusters within our horizon could accelerate to even higher rigidities \citep{Kourkchi17}.

Much of the infalling gas will have been processed by stars inside galaxies and either expelled as a wind or stripped by ram pressure. It is likely to start with a magnetic field of strength intermediate between that of the general IGM and what is required closer to the shock, say $10\,{\rm nG}$. In order for the magnetic field to grow to $\sim\mu{\rm G}$, it is necessary that the highest rigidity, escaping cosmic rays, far ahead of the shock, drive the growth of resonant Alfv\'en waves with wavevectors $k\sim r_L(R_{\rm max})^{-1}$ at a maximal rate ({\bf Sidebar Bootstrap mechanism}). 

\subsubsection{Filament accretion shocks}
A general feature of the large scale structure that develops in the expanding universe is the appearance of filaments which often appear as rod-like structures connecting rich clusters. They can be surrounded by cylindrical accretion shocks of modest strength. They are unlikely candidates for EECR acceleration but could contribute many of the $\sim\,{\rm EeV}$ UHECR \citep{Simeon15}. It is unclear whether or not our Galaxy lies within a filament but, if it does, then the cosmic rays transmitted downstream could contribute to the softer component of the cosmic rays we observe around and below the ankle \citep{Simeon15}. It is also possible that UHECR propagate along filaments \citep{Kim19}. 

\begin{figure}[h]
\includegraphics[width=5.7in]{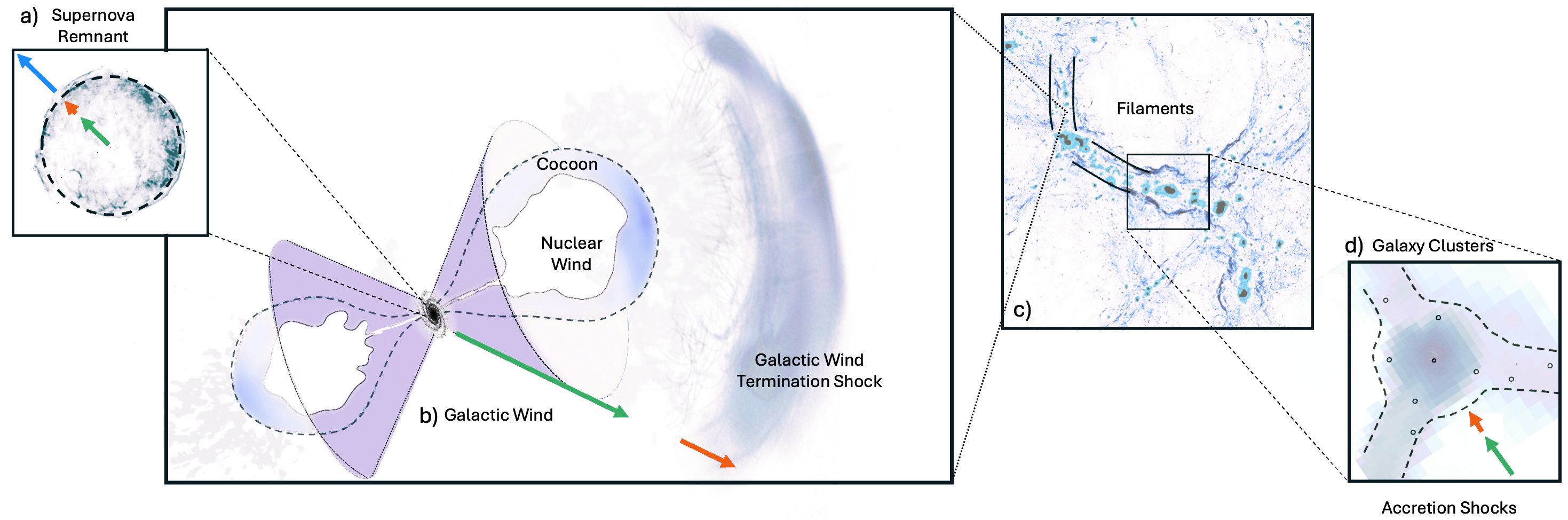}
\caption{Schematic of a slow acceleration hierarchical model. 
a) In this example, relying on DSA, suprathermal particles in the interstellar medium are first accelerated by expanding supernova-driven shock fronts. The blue velocity vector corresponds to the speed of the strong external shock through a stationary interstellar medium. The post-shock velocity in red is smaller so scattering centers on either side of the shock approach each other. The green vector represents the outflow from the explosion which is running into the post-shock flow, passing through a reverse shock. The outer shock expands quasi-spherically which allows the highest energy particles to escape upstream and dominate the observed Galactic cosmic ray spectrum. 
b) Galactic cosmic rays with energies up to $\sim1\,{\rm PeV}$ must escape the Galactic disk in $\lesssim$ 30~Myr. It is conjectured that they do this by driving a hydromagnetic wind, assisted by the hot gas in the interstellar medium. They do not return to the disk. In other galaxies, there may be a more powerful wind associated with their nucleus. This wind must have a large enough velocity (green arrow) to escape the gravitational potential well formed by the dark matter. At the outer radius of the Galaxy ($\sim200\,{\rm kpc}$), the wind passes through a quasi-stationary, internal termination shock.  The post shock velocity (red vector) is slower and if the shock is strong enough, there may be additional acceleration, contributing to the ``shin'' ($1\,{\rm PeV}\lesssim E\lesssim100\,{\rm PeV}$). This post-shock flow should be fast enough to allow the gas to escape and enter the CGM. The wind will likely flow outward through infalling gas clouds and satellite galaxies. It is the cosmic rays that are transmitted downstream that enter the CGM and IGM. The particles that escape upstream of the termination shock can be trapped within the halo and subject to additional acceleration. 
c) The cosmic rays that escape galaxies will suffer little loss beyond that associated with the expansion of the universe. and will accumulate in the intergalactic medium. Some of them will fall into quasi-cylindical shocks linking clusters. If we lie at the center of such a filament, the highest energy particles that are transmitted downstream may contribute to the ankle in our cosmic ray spectrum. d) Finally, a small fraction of the highest energy intergalactic cosmic rays may be convected by the intergalactic gas onto the cluster and are injected into a strong accretion shock. The upstream velocity (blue arrow) is much greater that the downstream velocity (red arrow). Roughly half of the highest energy cosmic rays accelerated, with $E\gtrsim100\,{\rm EeV}$, escape upstream, again facilitated by the spherical divergence. The acceleration and journey to Earth must be rapid enough to avoid losses.}
\label{h_dsa}
\end{figure}

\subsubsection{Galactic wind termination shocks and starburst galaxies}
Stellar and nuclear activity in galaxies can drive outflows which can create expanding shock fronts which will come to rest when their momentum density becomes comparable with the circumgalactic pressure \citep{veilleux05,thompson24}. Although such outflows are conspicuously absent from simulations there are observational reasons to take them seriously. For example, in the case of our Galaxy, the ratio of secondary elements like Li, Be, B to primary elements like C, N, O implies that Galactic cosmic rays have to leave the solar neighborhood in $\sim10\,{\rm Myr}$ and not return. Observations of nearby galaxies, such as NGC 253, clearly show such flows probably driven magnetocentrifugally although gas and radiation pressure are also important \citep{heesen11}. The terminal shocks may be very strong and are promising sites for particle acceleration. Typically they will be located beyond the virial radius associated with the dark matter potential well in which the galaxy sits, radii $\sim300\,{\rm kpc}$, and the characteristic speeds, $\gtrsim1000\,{\rm km\,s}^{-1}$, are typically several times the central escape velocity. The termination shocks are inwardly curved which implies that particles of all energies will be transmitted into intergalactic space and they are sufficiently strong that they can plausibly meet the luminosity challenge. However, these winds are individually less powerful than cluster accretion flows and so they would seem less likely to meet the rigidity challenge of EECR. However, they are quite plausibly sources for lower energy particles \citep{bustard17}.

Much attention has recently been directed towards starburst galaxies, epitomized by M82 and Cen A, on account to their reported association with the arrival directions of UHECR. The observed flows are relatively slow ($\sim100\,{\rm km\,s}^{-1}$) and low power ($\sim10^{41}\,{\rm erg\,s}^{-1}$). However, it has been conjectured that they are accompanied by faster and more powerful outflows including Ultra Fast Outflows from AGN, with near relativistic speeds which might meet the the rigidity challenge of UHECR but probably not the luminosity challenge \citep{beirao15,anchordoqui18,reeves20,peretti23}.

\subsubsection{Hierarchical Models}
In many DSA models, the accelerated particles are supposed to be a small minority of modestly suprathermal ions that separate from the background gas and participate as individual particles in the acceleration process. An alternative possibility is that the injected particles are already cosmic rays and that these are being re-accelerated by larger and more powerful shocks \citep{Blandford23}. This is especially relevant for cluster accretion shocks where the inflowing IGM is likely to contain a large PeV--EeV cosmic rays component accelerated in surrounding galaxies and expelled by their winds \citep{lacki15}.

\begin{textbox}[ht]
\section{BOOTSTRAP MECHANISM}
It is not known if a cosmic ray-driven strong turbulence can be maximal and self-sustaining as is required to meet the rigidity challenge. One way in which this can happen is if the highest rigidity cosmic rays, the only ones that can escape upstream, drive the resonant growth of Alfv\'en wavelengths as the rms  magnetic field will decrease upstream. These waves source a turbulence spectrum of shorter wavelength disturbances which are convected towards the shock front, super-Alfv\'enically, to meet and scatter successively lower rigidity particles as they reach their shorter diffusion lengths ahead of the shock. This is the essence of the ``magnetic bootstrap'' \citep{2007AIPC..921...62B}. It is not known whether or not a configuration like this will be self-sustaining. If so, it is quite likely that the cosmic ray - magnetic turbulence density undergo a limit cycle-type variation.  
\end{textbox}

We can generalize this idea and suppose that there is a hierarchy of acceleration processes \citep{Simeon23,simeon25} extending down to $\sim$ MeV energies where the output at one stage becomes the input at the next. In the simplest scheme, most of the acceleration is associated with shocks (Fig.~\ref{h_dsa}). The $\sim$GeV--PeV cosmic rays are accelerated by supernova remnant shocks, and escape the galaxy in a wind that they help drive. The highest-energy Galactic cosmic rays are injected at our Galactic Wind Termination Shock and further boosted to higher energies at this shock. We observe the returning cosmic rays upstream of the shock, and, more generally, the full spectrum is then transmitted downstream into the CGM by more powerful winds.
In this way, galaxies around clusters can supply the PeV--EeV cosmic rays to filaments and cluster accretion shocks that become the UHECR.
The contribution of PeV--EeV cosmic rays from galaxies around clusters could address injection, composition, and spectral challenges. 

\subsection{Relativistic Acceleration in Transient or Variable Sources}
\subsubsection{General considerations}
A very different class of accelerators is powered by compact objects within galaxies. Neutron stars and black holes are unavoidable endpoints of stellar evolution. Our galaxy probably hosts nearly a billion neutron stars and millions of black holes. They release their gravitational energy as neutrinos and expanding shells of gas that create the supernova remnants, bounded by the strong interstellar shock waves. These ultimately convert $\sim10^{-4}\,/\,10^{-5}$ of the gravitational energy released by individual neutron stars / black holes into Galactic cosmic rays. 

There are two other pathways to high energy particles, involving rotation. Conservation of angular momentum implies that newly formed compact objects are likely to be born spinning rapidly, perhaps close to their physical limits. Perhaps not coincidentally, large magnetic field is also likely to be present, along with MeV neutrinos and gas. A minority of core-collapse supernovae produce GRB, lasting seconds to minutes, perhaps the birth cries of stellar black holes. (GRB are also associated with rarer, binary neutron star mergers.) GRB produce highly-collimated, ultrarelativistic jets, which bore their way out through the collapsing stellar envelope, to drive beamed, relativistic blast waves into the surrounding interstellar medium, which can be observed for many months as ``afterglows''. These jets provide a possible source of UHECR, accelerated by relativistic shock waves (Section~\ref{bhjets}); the spinning black holes and neutron stars provide a slower release of energy, mediated by unipolar induction ({\bf Sidebar Unipolar Induction - Neutron Stars}), which have also been considered in this context. 

\begin{textbox}[ht]
\section{UNIPOLAR INDUCTION --- NEUTRON STARS}
A neutron star, radius $R$, endowed with a surface magnetic field ${\bf B}_0$ and rotating with angular velocity $\bf\Omega$ will generate an EMF. This is because it is an excellent electrical conductor and the electric field $\bf E$ must vanish in a local inertial frame rotating with the star. This, in turn, implies that, at a point near the surface of the star with position vector ${\bf r}$, the electric field ${\bf E}=-({\bf\Omega}\times{\bf r})\times{\bf B}_0$ in the non-rotating, inertial frame. If we specialize to a magnetic field that is stationary and axisymmetric with respect to $\bf\Omega$,the tangential component of this electric field must be continuous at the stellar surface and we immediately find that the potential difference across the field lines from the axis of symmetry can be expressed as $V\sim\Omega\Phi/2\pi$ where $\Phi$ is the magnetic flux. This is so large that charges will either be extracted   from the surface of the star or created {\sl in situ} at spark gaps. This, in turn, implies that ${\bf E}\cdot{\bf B}\sim0$ in the magnetosphere above the surface of the star and $V$ should be maintained across the magnetic field lines out to and beyond the ``light cylinder'' where $|{\bf\Omega}\times{\bf r}|=c$. The magnetic flux, $\Phi\sim B_0\Omega R^3/c$, that crosses the light cylinder is ``open''. There will also be a poloidal electrical current flowing along the magnetic field, $I\sim V/Q_{\rm eff}$, which is responsible for a toroidal magnetic field, comparable with the poloidal magnetic field. This allows us to estimate $Q_{\rm eff}\sim30\,{\rm ohm}$ outside the star. Electromagnetic energy flows inward and outward along the magnetic field and crosses the field lines, without dissipation, within the star, exerting a torque and extracting its rotational energy at a rate $L_{\rm EM}\sim V^2/Q_{\rm eff}$. The circuit completes and this power is dissipated beyond the light cylinder leading to particle acceleration, radiation and acceleration of outflowing plasma. Observed neutron stars are non-axisymmetric and can be non-stationary, with complex surface magnetic field, but these estimates suffice for our purpose.
\end{textbox}

Most normal galaxies have a massive ($\sim10^{6-10}\,{\rm M}_\odot$) black hole in their nuclei. These, too, are generally spinning and magnetized and can release their gravitational energy over billions of years by accreting gas through disks. This accounts for the prodigious powers of active galactic nuclei. They also release their rotational energy as relativistic jets which can escape the host galaxies and provide another possible source of UHECR. This is believed to be effected by a general relativistic version of unipolar induction  ({\bf Sidebar Unipolar Induction - Black Holes}). From a cosmological perspective, AGN jets are also transient, being most continuously active for times less than the UHECR propagation and loss times. 

Most of these transient sources are very powerful and can satisfy the luminosity challenge and the rigidity challenge insofar as the acceleration goes. The problems with these explanations are that the UHECR may be subject to catastrophic losses as they escape from their intense baryon and photon density environments, and that their rate should be sufficient within our horizon.

\subsubsection{Magnetars}
The simplest and best-observed type of relativistic accelerator is a neutron star. Regular pulsars have surface magnetic fields $B_0\sim10^{12}\,{\rm G}$ but only generate comparatively modest potential difference. For the Crab pulsar $V\sim30\,{\rm PV}$. Likewise for observed millisecond pulsars, with periods $P\sim1-10\,{\rm ms}$, but $B_0\lesssim10^9\,{\rm G}$ and observed magnetars with $B_0\sim10^{15}\,{\rm G}$, $P\gtrsim3\,{\rm s}$. These are not the source of UHECR.

However, magnetars are thought to comprise perhaps ten percent of newly-born neutron stars and it is reasonable to suppose that they are formed spinning rapidly. In this case, their rotational energy exceeds their magnetic energy, $\sim10^{49}\,{\rm erg}$, when $P\lesssim30\,{\rm ms}$, which requires them to be younger than $\sim1\,{\rm d}$. If we estimate a magnetar birth rate of $\sim10^{-3}\,{\rm yr}^{-1}$ per galaxy, then the magnetar luminosity density is ${\cal L}_{\rm magnetar}\sim3\times10^{36}(P/30\,{\rm ms})^{-2}\,{\rm erg\,s}^{-1}\, {\rm Mpc}^{-3}$, comparable with the estimated UHECR luminosity density. Magnetars could supply heavy nuclei, though neutron star surfaces may mostly contain H and He. They can satisfy the luminosity challenge, if they are born spinning faster than the Crab pulsar \citep{blasi00,arons03,Lemoine15,piro16}. They can function as unipolar inductors with $V\sim30(t/1\,{\rm d})^{-1}\,{\rm EV}$. Even greater acceleration is possible if the particles phase-ride the outgoing wave field in a manner similar to wake-field acceleration in the laboratory. Alternatively, magnetic energy can be made available explosively and intermittently over the lifetime of the magnetar. The brightest magnetar explosion, SGR~1806-20, released $\sim10^{47}\,{\rm erg}$ in $\sim0.1\,{\rm s}$ allowing $R_{\rm EM}\sim1000\,{\rm EV}$ (Eq~\ref{eq1}), meeting the rigidity challenge. However, meeting the full rigidity challenge is only possible if the acceleration occurs far enough away from  a nascent neutron star that radiative and expansion loss in the expanding outflow can be avoided, especially for heavy nuclei \citep{fang12,fang19}.

\begin{figure*}[]
\includegraphics[width=6in]{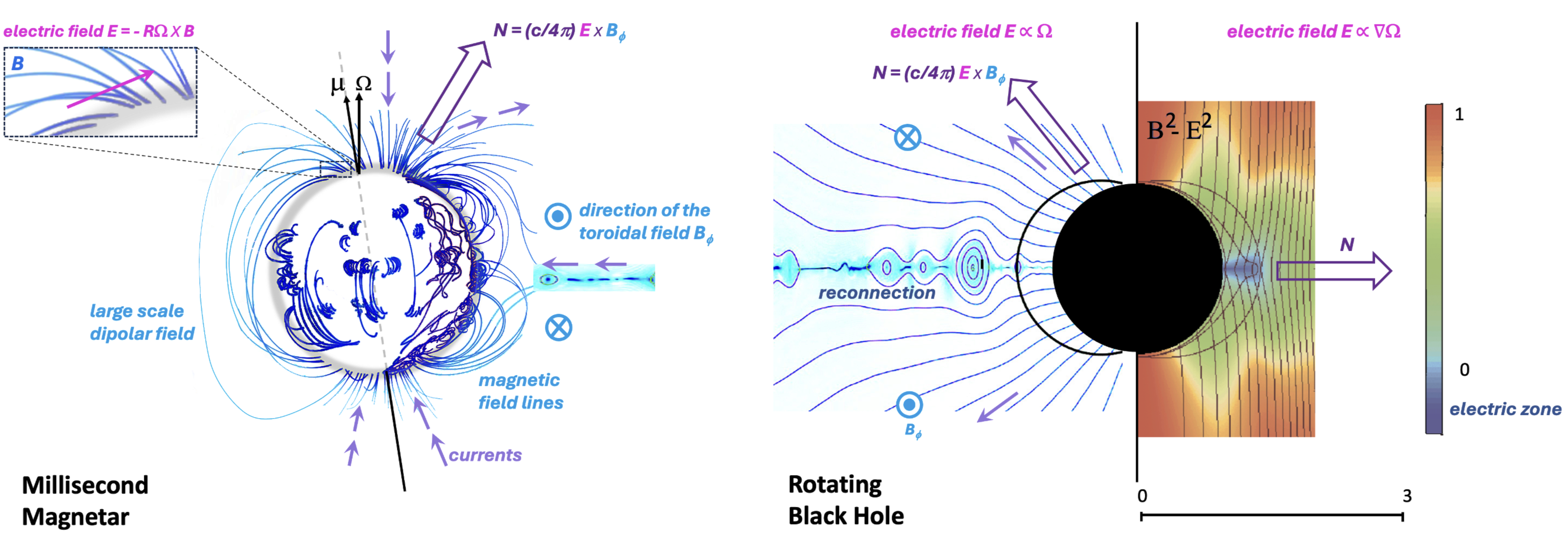}
\caption{Unipolar induction mechanism. \textit{Left}: structure of the magnetosphere of a millisecond magnetar. \textit{Right}: The geometry of magnetic fields around black holes is unknown, the left side show a split monopole configuration leading to an efficient EMF along the field lines \citep{BZ77} but not in the equator where a current sheet dissipates it. In this case UHECR acceleration would be similar to that of a magnetar (through direct one shot acceleration). The right side shows the ``ergomagnetosphere'' configuration \citep{BG22} with electric zones in the equator as shown by the colors \citep{Komissarov04} leading to an energy flux across the field lines which can then power winds that will collimates jets through strong oblique collimation shocks which can accelerate UHECR through DSA. }
\label{fig:unipolar}
\end{figure*}

\begin{textbox}[h!]
\section{UNIPOLAR INDUCTION --- BLACK HOLES}
Black holes threaded by magnetic field can also exhibit general relativistic unipolar induction, though there are some important differences from neutron stars. The effective resistance of a black hole is $\sim30\,{\rm ohm}$ but the magnetic flux must be held in place by gas orbiting outside the horizon. There is an internal resistance in the electrical circuit, within the event horizon. 

In the optimal case where the magnetosphere angular velocity is half that associated with the black hole, only $\sim$ half of the rotational energy associated with the black hole will be extracted electromagnetically with the remainder increasing the ``irreducible'' mass of the black hole. Most of the power extracted is in the form of antiparallel jets initially comprising electromagnetic Poynting flux, ${\bf N}=(c/4\pi){\bf E} \times {\bf B}$, associated with a large scale current flow (Fig.~\ref{fig:unipolar}). As the electromagnetic field is quite ordered, close to the source, a charged particle can cross the cross the jet and sample the full potential difference. This potential difference may be maintained in the outer parts of the jet, far from the nucleus, where catastrophic, radiative losses would occur. A simple estimate based upon an idealized, cylindrical jet containing a uniform, axial current density and similar to that used for neutron stars, leads to $Q_{\rm eff}\sim60\,{\rm ohm}$. 
\end{textbox}

\subsubsection{Black holes and their jets}\label{bhjets}
While the formation of a millisecond magnetar is sometimes associated with a minority of GRB, most GRB are thought to be associated with black hole formation, either directly during a core-collapse supernova or, after some delay, following the collapse of a magnetar. The time scale for gamma-ray energy release is seconds to minutes, followed by an X-ray to radio afterglow. Models for the formation and powering of these jets are varied but the most powerful GRB, with isotropic equivalent luminosities $>10^{52}$~erg s$^{-1}$, seem to require electromagnetic extraction of black hole rotational energy \citep{BZ77} ({\bf Sidebar unipolar induction}).  

It is not known whether the jet remains in an essentially electromagnetic, current-carrying, form all the way to its ends \citep[e.g.][]{blandford12,mbarek19} or if it converts to a baryonic form close to the source. Either way the jet can be an efficient accelerator of UHECR. In the electromagnetic limit, some current closure in the jet associated with the highest energy particles ($Q_{\rm eff}$) represents efficient particle acceleration \citep[c.f.][]{bell92}. In the purely baryonic limit, particle acceleration requires the formation of a shock. If this shock is mildly relativistic in its frame and it creates maximal magnetic turbulence ahead of the shock, the highest rigidities  reachable are essentially similar to what  you would have got  had it remained electromagnetic. 

For a $\sim3\,{\rm s}$ GRB the jet will have a length $\sim10^{11}\,{\rm cm}$ in the source frame, comparable with the photospheric radius of the collapsing star. In the frame moving out with the jet, the length is $\Gamma$ times longer. After the jet emerges from the star into the circumstellar medium,  it will have the form of an expanding cylinder moving ultrarelativistically along its axis with Lorentz factors of typically hundreds in a radial direction and surrounded by shocked, ambient gas. If UHECR are accelerated by GRB jets, this must happen in the interval between their escape radius, estimated to be very roughly $10^{15}\,{\rm cm}$ and the radius where the jet decelerates and transfers its energy to the surrounding gas, perhaps $\sim10^{17}\,{\rm cm}$.  

Variation in the jet speed, due to the black hole source, will cause internal shocks to form within the jet. Deceleration of the jet by the swept up gas will cause a reverse shock. Interaction with the surroundings can lead to recollimation shocks and instabilities. All of this creates an accompanying, complex network of shocks. From our discussion of DSA, we see that these shocks are most likely to meet the rigidity challenge when they are mildly relativistic. In this case, the Larmor radius for $R'_{\rm EM}$ in the post-shock gas in the frame of the shock is of order the shock width. Now the jet power, $L'_{\rm EM}$  entering such a shock in its frame is $\sim\Gamma^{-2}$ times $L_{\rm EM}$, the jet power in the source frame. So, very roughly, $R'_{\rm EM}\sim(Q_{\rm eff}L'_{\rm EM})^{1/2}$ where $Q_{\rm eff}\sim30\,{\rm ohm}$. Transforming into the source frame, $R_{\rm EM}\sim(Q_{\rm eff}L_{\rm EM})^{1/2}$. This is many orders of magnitude larger than the rigidity required for EECR, $\sim10\,{\rm EV}$.

\begin{figure*}
\includegraphics[width=5.7in]{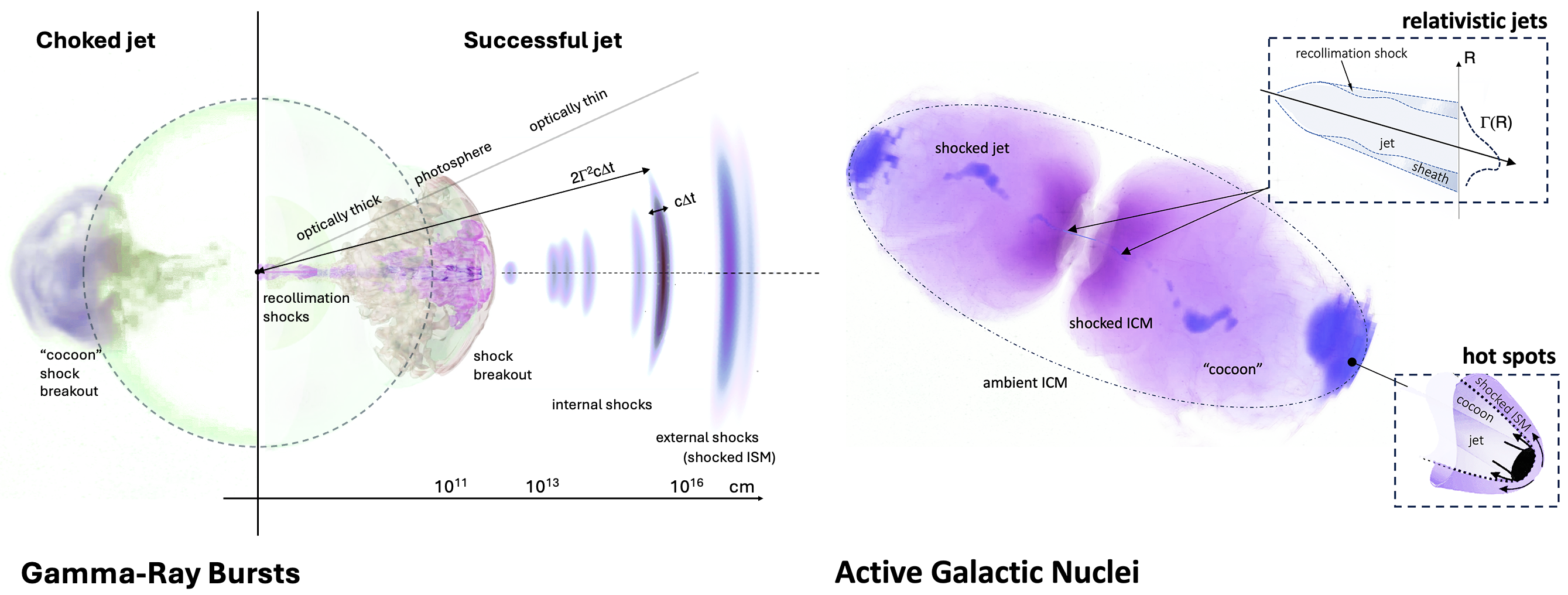}
\caption{{\it Left}: Potential locales for UHECR acceleration in GRBs jets include internal shocks   (believed to be the emission sites of the prompt X-rays and gamma-rays), external reverse shocks (believed to be the emission sites of optical flash and radio flare), and external forward shocks (believed to be the emission sites of the X-ray to radio  afterglow). Mildly relativistic internal and external reverse shocks are the best candidates sites for UHECR acceleration as they occur in the jet comoving frame and the escaping UHECR gets a factor hundred boost in their energy. The external forward shock is disfavored due to its ultrarelativistic velocity, unless a strong ordered field is present. In the optically thick regime, internal (recollimation) shocks may lead to neutrino production both in choked and successful jet cases. The mildly relativistic shock breakout and external shock can also be a source of PeV--EeV cosmic rays. {\it Right}: Potential locales for UHECR acceleration in AGNs jets include recollimation shocks, internal shocks and external shocks (hot spots).}
\label{jets}
\end{figure*}

Now let's consider the DSA mechanism.  UHECR could be accelerated either through the external ultrarelativistic shock \citep{vietri95} or through the internal shocks \citep{waxman95}. Internal shocks are mildly relativistic and therefore can account for the highest rigidities.
It has been shown that the external, ultrarelativistic shocks are very inferior accelerators because relativistic kinematics makes returning particles to a shock front from downstream too improbable \citep{Gallant99,niemiec06}.

There is, however, a problem with the luminosity challenge. The high luminosity GRB rate is very roughly $\sim10^{-5}\,{\rm yr}^{-1}$ per galaxy and even if we associate a near maximal electromagnetic energy $\sim10^{52}\,{\rm erg}$ with each GRB, the resulting luminosity density is only ${\cal L}_{\rm EM}\sim3\times10^{37}\,{\rm erg\,s}^{-1}\,{\rm Mpc}^{-3}$, almost equal to what is required.  
Furthermore, this power is nearly two orders of magnitude more than the gamma-ray luminosity density \citep{Eichler10, Globus2015a} and so, if the high luminosity GRB really are the source of UHECR then, perhaps, they should be renamed CRB!

Massive black hole jets exhibit many of the same qualitative features as GRB jets. Their powers range up to $\sim10^{44}\,{\rm erg\,s}^{-1}$ and the Lorentz factors start with $\Gamma\sim10$, smaller than for GRB. The most powerful AGN jets could just meet the rigidity challenge for mildly relativistic DSA. However they are much less common locally than at peak activity, around a redshift $z\sim2$. 
However the radio sky may have looked different several hundred million years in the past. Centaurus A, Virgo A and Fornax A may have been more active in the past as a result of galaxy mergers and may have contributed to the anisotropy \citep{Harari14}. 

AGN jets have many advantages. They naturally form mildly relativistic shocks which are optimal accelerators \citep{niemiec06}. They can be even more effective if they remain essentially electromagnetic. They might be seen as individual sources \citep{2018JCAP...02..036E}, as has been claimed for Cen A \citep{Rieger09} and they might account for the anisotropy. (For Cen A to be the dominant source of UHECRs, the local EGMF would have to be strong, $\gtrsim$80~nG \citep{Farrar13}). Acceleration of the UHECRs in the backflow \citep{Wykes13,Matthews18} and at the termination shock \citep{2023A&A...676A..23C} have shown to be efficient.

\begin{textbox}[ht]
\section{RELATIVISTIC RECONNECTION}
Nonrelativistic magnetic reconnection occurs because the magnetic field in conducting plasma often has a strong gradient. The resulting current layers naturally become thinner and unstable. Even when the resistivity is small, it is possible to create isolated sites where magnetic flux is no longer ``frozen'' into the underlying plasma and the field lines exchange partners at a rate controlled by the effective resistivity and release energy, some of which channeled into particle acceleration \citep{sironi14}. While non-relativistic reconnection is relatively inefficient, relativistic reconnection is found to be both efficient and capable of accelerating to high rigidity. In practice, the geometry is complex with relativistically moving plasmoids creating additional particle acceleration. Typically, the electric field in the reconnection site is $\sim0.1$ times the magnetic field strength. Associating $L_{\rm EM}$ with the Poynting flux incident upon a single reconnection site, leads to $Q_{\rm eff}\sim10\,{\rm ohm}$.
\end{textbox}

Observed AGN jets are often delineated, at radio wavelengths, by emission from their surfaces, or sheaths. Sheaths can be interpreted physically as either return current sheets or fluid boundary layers, most likely both. UHECR could be accelerated here instead of at transverse shocks. The physical process that has been most invoked is relativistic reconnection ({\bf Sidebar Relativistic Reconnection}) \citep{Giannos10} or ``one shot'' acceleration \citep{2015ApJ...811L..38C}. Again there is a minimum radius where the radiation density is low enough to allow the UHECR to escape. At this radius, an individual reconnection region will be comparatively small and the electromagnetic luminosity too small to meet the rigidity challenge.  

Another possibility, that could operate within an electromagnetic jet, is that there is a major rearrangement of large flux tubes without significant reconnection, a process called magnetoluminescence ({\bf Sidebar Magnetoluminescence}). Acceleration of EECR requires an electromagnetic energy $\gtrsim10^{45}\,{\rm erg\,s}^{-1}$ which is a challenge in the local sources we observe.

There are other ways AGN have been proposed as UHECR sources and could meet the rigidity challenge. These include short-lived, though highly luminous  TDE, which occur when a normal star is nudged so close to a massive black hole that it is ripped apart. Hot spots have been associated with TDE in nearby galaxies \citep{pfeffer17}. A minority of these relatively common events form radio jets, often with some delay, which could  be sources of UHECR \citep{Farrar14,2023JCAP...11..049P}. The jets could carry as much as $\sim10^{45}\,{\rm erg\,s}^{-1}$. Another type of radio source, known as a Compact Symmetric Object may also be TDE \citep{Readhead94}. Together, these sources can satisfy the luminosity challenge. 

\begin{textbox}[ht]
\section{MAGNETOLUMINESCENCE}
Magnetic field has a propensity to organize itself into restless, tangled  flux tubes. Individually, these will be twisted and are often resolvable into smaller tubes that can stray into the surrounding medium forming a rough surface, rather similar to sisal rope. Pursuing this analogy can lead to individual flux tubes forming ``slip knots''. As the tubes are under tension, they are likely to resolve themselves, releasing a lot of magnetic energy at the speed of light \citep{blandford17,yuan16} so that the electric field is of order the magnetic field strength. As this occurs over a large volume the resistance is likely to be quite large and we estimate $Q_{\rm eff}\sim100\,{\rm ohm}$.
\end{textbox}

\section{DISCUSSION}
\begin{extract}
``The greatest pleasure a scientist can experience is
to encounter an unexpected discovery'' -- James Cronin
\end{extract}

The resolution of the long standing mystery of the origin of UHECR will require an interdisciplinary approach on three complementary fronts: particle physics, cosmic-ray physics and astrophysics. We began this review by asking three entangled questions. Let us return to them and summarize progress since the last Annual Review on this topic \citep{2011ARA&A..49..119K} including recent developments.

$\bullet$ ``What are they?''
Thanks, in part, to major efforts directed towards reconciling and improving the calibration of the two major facilities, Auger and TA, there is strong evidence that UHECR are a mixed composition of atomic nuclei with the composition becoming heavier moving from the ankle at $\sim5\,{\rm EeV}$ to $\sim300\,{\rm EeV}$. We have today seven state-of-the-art QCD models that are used for the air showers simulations, and there are still significant differences between them leading to different interpretation of the UHECR composition. All models  show a significant muon deficit with respect to measurements and effort have been made during the recent years to address what is now called the muon puzzle.

The most important theoretical development would be a better understanding of the hadronic physics at the highest energies. This is crucial to determine if the muon puzzle is an experimental misunderstanding or new physics. If the latter, the hadronic models currently used would need to be improved.  Future data from oxygen beam collisions at the LHC,  \citep[as proposed in][]{Citron19} will be crucial to understand the difference between proton-proton and nucleus-nucleus interactions. Oxygen-proton collisions are representative of the first interaction in an air shower.

$\bullet$ ``How do we detect them?''
TA is being upgraded with the goal of quadrupling its geometrical exposure, which would make it comparable with present Auger \citep{TelescopeArray:2021}. Meanwhile, Auger is upgrading its surface detectors, adding radio detectors and underground muon detectors which should help address the muon puzzle \citep{Stasielak22}. We made progress in understanding radio signal associated with air showers (in particular with LOFAR) and the construction of the world-largest radio detector (AugerPrime) has started. The signals from all Auger detectors will be processed simultaneously and provide better data quality for air shower modeling.

An important observational development would be to be able to determine the composition of primary UHECR on an event-by-event basis. New methods for identifying the mass composition of UHECR using deep learning have been introduced by various working groups and applied to the data from TA and Auger \citep{Fujii:2024sys}. New detection concept have been proposed and would allow to probe possible different composition-dependent observables. On the anisotropy front, a full sky observatory with larger exposure is necessary to accumulate sufficient statistics at the highest energies. 

There are new missions being developed. A space-based mission, the Probe of Extreme Multi-Messenger Astrophysics, POEMMA, has similar goals and will also be able to observe earth-skimming, PeV neutrinos \citep{POEMMA}. The JEM-EUSO program, Joint Exploratory Missions for Extreme Universe Space Observatory, is the realization of a space mission devoted to look down on the Earth’s atmosphere for air fluorescence flashes from EAS \citep{Coleman2023}. After three balloon flights (EUSO-Balloon, EUSO-SPB1, and EUSO-SPB2) and a mission in the ISS (MINI-EUSO), another balloon flight is in preparation, the POEMMA Balloon with Radio, PBR \citep{2024NIMPA106969819B}. As its name suggests, in addition to combining the fluorescence camera and the Cherenkov camera in the same telescope, it includes radio shower detection. A new and important target of PBR concerns High-Altitude Horizontal Air Showers.
 On the longer timescale, there is a proposal to construct a Global Cosmic-ray Observatory, GCOS, which should achieve this resolution for $10\lesssim E\lesssim50\,{\rm EeV}$ as well as improve the angular resolution to better than $\sim30\,{\rm arcmin}$ \citep{Batista23}.    (An interesting, though largely unexplored, possibility is to observe the doubly-dark lunar surface using orbiting spacecraft to seek radio \citep{RomeroWolf20} or optical  flashes from UHECR impacts.)

$\bullet$ ``What is their origin?'' 
Some source models (in particular, top-down scenarios) are currently disfavored by direct cosmic-ray and indirect gamma-ray observation. No new source classes have emerged that meet our challenges and we seem to be left with relativistic jets and intergalactic shock fronts for the EECR. The exclusions are not hard and lower power sources including  starburst winds, tidal disruption events could account for some of the $\sim1\,{\rm EeV}$ UHECR cosmic rays. A framework to explore the different signatures from continuous \citep{Harari14} or transient \citep{Harari21} sources. Indirect observation, through multi-messenger astronomy \citep{Murase19} may help to constrain the source properties. Neutrinos are signature of cosmic-ray sources but there are still no secure ($\geq5\sigma$) identifications with a class of astronomical sources. It should be emphasized that due to the delay induced by GMF and EGMF secondary neutrinos, photons, and gravitational waves should not be detected in time coincidence with UHECRs if the sources are not continuously emitting particles, but are transient.

The most important astrophysical development would be a solid framework for UHECR astronomy. This needs progress on two fronts: first, a better determination of the composition, on an event-by-event basis, and second,  a more confident and well-validated understanding of the magnetic field within our Galaxy and its neighbors. If we are successful in doing this, we will be able to constrain the source(s) of UHECR by focusing on the subset of UHECR that experience minimal deflections by the GMF. In other words we will be able to perform  a ``tomography'' of the sources in different mass and energy groups. For example, the GZK horizon for EECR is limited to the Local Group for CNO nuclei and $\sim40\,{\rm Mpc}$ for heavy nuclei. Furthermore, the event-by-event rigidity information would allow to identify transient sources, which have an unique temporal ordering in rigidity of the events coming from a same cosmic-ray burst \citep{Farrar08,2023ApJ...945...12G}. The observational situation should improve dramatically in the next few years using a million compact extragalactic radio sources \citep{Gaensler24}. Much attention is being paid to developing a better understanding of the turbulent component of the magnetic field through observation and simulation.

\begin{issues}[FUTURE ISSUES AND DIRECTIONS]
\begin{enumerate}
\item{The application of more varied shower diagnostics, better interaction models from the LHC and machine learning should help us make more accurate descriptions of the primary cosmic rays.}
\item{The goal of practising reliable cosmic ray astronomy will require measuring the energies, the atomic numbers and the incoming directions associated with single events. It will also require a secure model of the magnetic field extending out into halo an the local group. This includes developing better models of MHD turbulence.}
\item{With the opening up of PeV neutrino, TeV gamma-ray as well as feV gravitational wave astronomy, we should be able determine which transient or variable sources have the capacity to meet the 10 EV rigidity challenge with the required efficiency. It may turn out that the UHECR accelerators are complementary to those that dissipate effectively, through the neutrino channel, at PeV energies.}
\item{Analytic, fluid and kinetic models of particle acceleration are still unable to account in detail for the apparent success of cosmic sources to meet the rigidity and luminosity challenges we have described. Much can be learned from better observed, less extreme accelerators. More emphasis on the phenomenology should help us understand where UHECR are produced and if they are injected there out of the thermal plasma  or are part of a hierarchical sequence.}
\item{Perhaps the greatest need is for larger facilities - such as GCOS on the ground and POEMMA in space - to boost the event rate considerably. This will allow us to provide secure answers to the questions we have posed and access the immense discovery space that is there to be explored. As we have tried to emphasize, there is both the opportunity to make secure source identifications and maybe ``unexpected discoveries'' in time for the next Annual Review on UHECR.}
\end{enumerate}
\end{issues}

\section*{DISCLOSURE STATEMENT}
The authors are not aware of any affiliations, memberships, funding, or financial holdings that
might be perceived as affecting the objectivity of this review. 

\section*{ACKNOWLEDGMENTS}
``Perhaps the most important observation of cosmic rays, that a theory of their origin must account for, is that they exist.'', our friend and colleague David Eichler wrote in 1986. The origin of cosmic rays was a life-long fascination of him and we dedicate this review to his memory. We express our gratitude to long-term collaborators who contributed some key work cited here: Bram Achterberg, Denis Allard, Glennys Farrar, Anatoli Fedynitch, Stefan Funk, Jerry Ostriker, Etienne Parizot, Tsvi Piran, Paul Simeon and helpful conversations with Tony Bell, Damiano Caprioli, Don Ellison, Ke Fang, Ralph Engel, Bryan Gaensler, Yannis Liodakis, Alexandre Marcowith, John Matthews, Payel Mukhopadhyay, Igor Moskalenko, Angela Olinto, Enrico Peretti, Troy Porter, Frank Rieger, Steve Reynolds, Lorenzo Sironi, Anatoly Spitkovsky, Hiroyuki Sagawa, Michael Unger, Alan A. Watson. We gratefully acknowledge the support of the Simons Foundation (MP-SCMPS-00001470, N.G., R.B.).
 
\noindent
\bibliographystyle{ar-style2}

\end{document}